\documentclass[12pt,preprint]{aastex}

\usepackage{epstopdf}

\begin{document}

\title{Orbital Parameters for the Two Young Binaries VSB 111 and VSB 126}

\author{N. Karnath\altaffilmark{1,2,3}, L. Prato\altaffilmark{1},
L. H. Wasserman\altaffilmark{1}, Guillermo Torres\altaffilmark{4}, B. A. Skiff\altaffilmark{1}, R. D. Mathieu\altaffilmark{5}}

\altaffiltext{1}{Lowell Observatory, 1400 West Mars Hill Road, Flagstaff,
AZ 86001, USA; nicole@lowell.edu, lprato@lowell.edu, lhw@lowell.edu, bas@lowell.edu}
\altaffiltext{2}{Department of Physics and Astronomy, Northern Arizona University, Flagstaff, AZ 86001, USA.}
\altaffiltext{3}{Department of Physics and Astronomy, University of Toledo, Toledo, OH 43606, USA.}
\altaffiltext{4}{Harvard-Smithsonian Center for Astrophysics, 60 Garden St., Cambridge, MA 02138, USA: gtorres@cfa.harvard.edu}
\altaffiltext{5}{Department of Astronomy, University of Wisconsin-Madison, Madison, WI 53706, USA; mathieu@astro.wisc.edu}

\begin{abstract}

We report orbital parameters for two low-mass, pre-main sequence,
double-lined spectroscopic binaries
VSB 111 and VSB 126. These systems were originally identified as single-lined
on the basis of visible-light observations.  We obtained high-resolution, infrared spectra
with the 10-m Keck II telescope, detected absorption lines of the secondary stars, and
measured radial velocities of both components in the systems. The visible light spectra were
obtained on the 1.5-m Wyeth reflector at the Oak Ridge Observatory, the 1.5-m Tillinghast
reflector at the F.\ L.\ Whipple Observatory, and the 4.5-m equivalent Multiple Mirror Telescope.
The combination of our visible and infrared observations of VSB 111 leads to a period of 902.1$\pm$0.9 days, an
eccentricity of 0.788$\pm$0.008, and a mass ratio of 0.52$\pm$0.05. VSB 126 has a period
of 12.9244$\pm$0.0002 days, an eccentricity of 0.18$\pm$0.02, and a mass ratio of
0.29$\pm$0.02. Visible-light photometry, using the 0.8-m telescope at Lowell
Observatory, provided rotation periods for the primary stars in both systems,
3.74$\pm$0.02 days for VSB 111 and 5.71$\pm$0.07 days for VSB 126.  Both binaries
are located in the young, active star-forming cluster NGC 2264 at a distance of $\sim$800 pc. 
The difference in the center-of-mass velocities of the two systems is consistent with the radial velocity gradient
seen across NGC 2264.  To test the evolutionary models for accuracy and consistency,
we compare the stellar properties derived from several sets of
theoretical calculations for pre-main sequence evolution with our dynamical results.

\end{abstract}


\section{Introduction}

A binary star provides one of the few ways to directly measure stellar masses. The mass is the most important property of a star and determines its basic structure and properties, as well as the duration of stages of evolution throughout its lifetime. A double-lined spectroscopic binary (SB2) enables the orbit to be solved from spectroscopically-determined radial velocities (RVs), yielding the orbital elements (except $i$) and mass ratio, $q$, for the component stars. Component masses may be found if the inclination of an SB2 can be measured in the case of an eclipsing binary system ($i$=90$\,^{\circ}$) or a resolved visual binary. 

The work presented here is part of a long-term program to measure young SB2 mass ratios and eventually component masses, $M{_1}$ and $M{_2}$ (Prato et al. 2002; Schaefer et al. 2012). 
Such results will be facilitated with the advent of
high-precision astrometric missions such as GAIA. Ultimately, a large sample of precise, dynamical masses will help to anchor models of pre-main sequence (PMS) evolution to a solid observational basis, particularly for systems with stellar
mass $<$1 M$_\odot$ (e.g., Hillenbrand \& White 2004; Simon 2008). Furthermore, precise measurements of the mass-ratio distribution provide critical input for theories of binary star formation (Bate 2009). Our present goal is to build up a large sample of young star mass ratios. 

Vasilevskis, Sanders, \& Balz (1965) included VSB 111 (V810 Mon) and VSB 126 (2MASS J06410777+0944030) in their study. Since then, several papers adopted the VSB acronym; therefore, we use those names throughout the paper. VSB 111 and VSB 126  are located in the young 
cluster NGC 2264. This cluster is an active region of star formation,
with subclusters of suspected members spread across several parsecs. Dahm (2008) published a summary of the papers on NGC 2264. 
There are numerous estimates of the distance to NGC 2264, ranging from 700 pc (Feldbrugge \& van Genderen 1991) 
to 950 pc (P\'{e}rez et al. 1987). 
We adopted a distance of 800 pc from Walker (1956) because it was a median value within the estimated distances published in Dahm (2008). We adopted an uncertainty of $\pm$100 pc to account for the large spread of published distances. Additionally, there is a large age range cited in the literature. Estimates for the age of the cluster range from 0.1 Myr (Flaccomio et al. 1999; Rebull et al. 2002) to 10.0 Myr (Flaccomio et al. 1999), 
but $\sim$3.0 Myr is common (Walker 1956; Mendoza \& G\'{o}mez 1980; Feldbrugge \& van Genderen 1991; Sung et al. 2004; Ramirez et al. 2004; Flaccomio et al. 2006).  This is not particularly surprising, however, as the ages of stars in young clusters typically appear to span a range of millions and
even tens of millions of years (e.g., Hillenbrand 1997).
 
Haisch et al. (2001) observed that about half of the stars within clusters lose their disks in approximately 3 Myr. Mathieu (1994) lists the H$\alpha$ equivalent widths for VSB 111 as -0.3 \AA \ and VSB 126 as -1.9 \AA, indicating a lack of any appreciable gas accretion in these systems.
WISE and Spitzer data show no indication of warm dust in VSB 111 and VSB 126. We therefore believe they are diskless.  The basic properties of the systems are listed in Table 1.

Mathieu (1994) originally derived single-lined spectroscopic (SB1) orbital solutions for VSB 111 and VSB 126. Given his mass functions for the systems, we knew a priori that the minimum mass ratios were both likely to be small.
When a low mass ratio binary system is observed in visible light, the primary star's
spectrum dominates the faint secondary signal because of the low flux ratio. Observing in the infrared (IR) allows for detection of the secondary component and thus solution
of the system as an SB2 because the flux ratio is more favorable in the long-wavelength
Rayleigh-Jeans regime. This approach was demonstrated in Prato (1998), 
Mazeh et al. (2002), Prato et al. (2002), and Mazeh et al. (2003).
Observing PMS binaries in the IR can also be advantageous for young stars obscured in
dusty star forming regions (SFRs).  

In this paper we contribute dynamical mass-ratio measurements for two young binaries, one short-period (several days) and one long-period (few years), to the as yet relatively small sample of $\sim$50 PMS SB2s. 
Furthermore, our results provide improved
orbital parameters for both binaries.  Based on the stellar characteristics and light curves
measured as part of this program, we also estimated the inclinations of the 
binary orbital planes and the stellar rotation axes.
In \S2 we describe our observations and data reduction. In \S3 we provide our analysis
and results. A discussion appears in \S4 and \S5 summarizes our findings.

\section{Observations and Data Reduction}

\subsection{Visible Light Spectroscopy}

Visible light spectroscopic observations of VSB 111 and VSB 126 were carried out at the Harvard-Smithsonian Center for Astrophysics (CfA) using three different telescopes equipped with nearly identical echelle spectrographs: the 1.5-m Wyeth reflector at the Oak Ridge Observatory (Harvard, Massachusetts), the 1.5-m Tillinghast reflector at the F.\ L.\ Whipple Observatory (Mount Hopkins, Arizona), and the 4.5-m equivalent Multiple Mirror Telescope (also on Mount Hopkins), prior to its conversion to a monolithic 6.5 mirror. A single echelle order 45\,\AA\ wide and centered near 5190\,\AA\ (including the \ion{Mg}{1}\,b triplet) was recorded with intensified photon-counting Reticon detectors at a resolving power of $\lambda/\Delta\lambda = 35,\!000$. We collected 69 spectra for VSB 111 between 1984 January and 1996 January, with signal-to-noise ratios between 8--20 per resolution element of 8.5 km s$^{-1}$. For VSB 126 we collected a total of 30 spectra from 1984 March to 1992 February, at signal to noise ratios of 5 to 11 per resolution element. These data are largely the same as those on which the results of Mathieu (1994) are based. The details of the observations are provided here for the first time. The specific dates of observation appear in the first column of Tables \ref{tbl:vsb111visiblerv} and \ref{tbl:vsb126visiblerv} . Spectra were reduced using standard IRAF procedures for echelle data (e.g., Torres et al. 1997). 

\subsection{Infrared Spectroscopy}

VSB 111 and VSB 126 were observed in the IR with the Keck II 10-m telescope on Mauna Kea.
The observations for VSB 111 and VSB 126 were made between 2001 December and 2012 January (Tables 4 and 5).
The Keck facility near-IR spectrograph, NIRSPEC, was used to obtain $H$-band data at a central wavelength of $\sim$1.555~$\mu$m (McLean et al. 1998; 2000).
 NIRSPEC employs a 1024 x 1024 ALADDIN InSb array detector.  The $0\farcs288$ (2 pixel) x $24''$ slit yielded a resolution of R=30,000. The slit viewing camera, SCAM, which uses a 256 x 256 HgCdTe detector with 0\farcs18 pixels, facilitated source acquisition and guiding. For VSB 111, integration times for the individual frames were 240 s or 300 s; two to eight frames were taken on each observing date.
For VSB 126, integration times for the individual frames were
300 s or 360 s; six to fourteen frames were taken on each visit.  For both systems, the
signal to noise ratio was always $>$100 and typically 200 or higher.
Background subtraction was achieved between consecutive spectra by nodding the telescope $12''$
to dither the target between two positions on the slit and using an A-B-B-A pattern of observation.

Our $H$-band setting (echelle$=63\fdg04$, cross-disperser$=36\fdg3$) has a central wavelength of $\sim$1.555~$\mu$m, which corresponds to order 49 (1.545$-$1.567~$\mu$m)
for NIRSPEC. We focused on order 49 because it is free of telluric absorption lines and contains well-spaced OH emission lines.
De-excitation of the hydroxyl molecule (OH) in the Earth's atmosphere produces these emission lines;
they are distributed throughout order 49 and were used to calibrate the wavelength zero-point and dispersion (Rousselot et al. 2000).  Order 49 is also rich in atomic and molecular species in the stellar photospheres that allow us to identify spectral features of both
warm and cool stars. 

NIRSPEC's optics optimize the throughput of photons but the resulting spectra do not fall along uniform rows; the light falls in curved swaths. Therefore, we needed to rectify the data spatially and spectrally. For all data reduction we used the REDSPEC software package, written at UCLA by S. Kim, L. Prato, and I. McLean, and created specifically for the analysis of NIRSPEC data. The lack of terrestrial absorption lines in order 49 eliminates the need to divide by telluric standard star spectra. Median images of 10 flat and 10 dark files were used to create a master flat and dark in order to correct for effects from the detector. 

The REDSPEC module spatmap was used to remap the raw images onto a uniform interval coordinate system in the spatial (cross-dispersion) direction by adding two nodded frames together and fitting a $3^{rd}$ order polynomial to each spectral trace. The specmap module was used to remap the raw images onto uniform interval coordinate system in the dispersion direction. A $2^{nd}$ order polynomial was fit to each OH line to identify the approximately vertical trace. Then, a $2^{nd}$ order polynomial fit in the dispersion direction was made to these wavelength locations for each row around the spectral trace. The final step of REDSPEC was to divide by the dark-subtracted master flat, remove any bad pixels and fringing, and clip out and sum the rows
containing the rectified spectra. All spectra were corrected for heliocentric motion and the Earth's rotation using JSkyCalc$^{1}$ \footnotetext[1]{http://www.dartmouth.edu/$\sim$physics/faculty/skycalc/flyer.html}and are shown in Figures 1 and 2 for VSB 111 and VSB 126, respectively.

\subsection{Visible Light Photometry}

The photometric data were taken on twelve nights between 2012 September 14 and October 11 using the Lowell 0.8-m telescope in robotic mode. A total of 45 images were obtained for VSB 111 and 44 images for VSB 126. The CCD camera houses
an e2v 2k$\times$2k chip providing a $15' \times15'$ field. The image scale is
$0\farcs91$/pixel when binned 2$\times$2, as here. A median bias frame was produced
every night from several dozen zero-integration time exposures. Twilight flats are scripted for each filter in use on a rotating schedule and taken when obtainable. Further details about the telescope and its conversion to robotic use can be found in Buie (2010). Pairs of 150 s exposures using a Johnson $V$ filter were taken at each visit to NGC 2264. These were observed among roughly a dozen other targets in the queue schedule whenever the field was at less than 2.0 airmasses.  Since the observing was done near the beginning of the season, typically four visits (up to eight) were made to the field each night roughly a half hour to a full hour apart. The images were reduced with the commercial photometry package $Canopus$ (Warner 2011). The software includes a photometric catalogue with $BVRI$ data derived from 2MASS $JHK_s$ photometry (Warner 2007), as well as more traditional published $BVRI$ photometry, and Sloan $griz$ catalogues.  These provide photometric zero-points and color
indices ($\sim$0.03 mag) for the entire sky via on-chip differential photometry without the need to observe primary standards. $Canopus$ plots the run of instrumental magnitudes versus its internal photometric catalogue of all stars in the image. The averaged photometry from the two frames at each visit to the field was used for analysis.

In the NGC 2264 region the majority of the brighter stars are identified or suspected variables. All five of the comparison stars we selected appear in the NSV catalogue or its supplements (Samus et al. 2012), and have been hitherto suspected to be variable with ranges of up to a half magnitude. Table \ref{tbl:compstars} lists the coordinates and aliases of the comparison stars. The five comparison stars were chosen to be similar in color to the two binaries, which helps minimize errors in the photometry
resulting from color terms and differential extinction. All five comparison stars are demonstrably constant at the half-percent level on the timescale of weeks. The reduced data for VSB 111
and VSB 126 are shown in Tables \ref{tbl:vsb111photometry} and \ref{tbl:vsb126photometry}, respectively. The light curves for the primary star in each system are shown in Figures 3 and 4 for VSB 111 and VSB 126, respectively.

\section{Analysis}

\subsection{Radial and Rotational Velocities}

All our visible light CfA spectra appear single-lined. The templates that best match the spectra implicitly provide an estimate of the spectroscopic parameters of the primary stars. RVs were obtained by cross-correlation using the IRAF task {\tt XCSAO}, with templates chosen from a large library of calculated spectra based on model atmospheres by R.\ L.\ Kurucz (Nordstr\"om et al. 1994; Latham et al. 2002).  
Of the four main parameters of these templates (effective temperature $T_{\rm eff}$, rotational velocity $v \sin i$, metallicity [Fe/H], and surface gravity $\log g$), the ones affecting the velocities the most are $T_{\rm eff}$ and $v \sin i$. Consequently, we held $\log g$ fixed at a value of 3.75 for both stars and assumed solar metallicity.  The optimum $T_{\rm eff}$ and $v \sin i$ values were determined by running grids of cross-correlations seeking the maximum of the correlation coefficient, averaged over all exposures (Torres et al. 2002).
We obtained $T_{\rm eff}$ = 5300$\pm$100$\,K$ and $v \sin i$ = 31$\pm$2 km s$^{-1}$ for VSB 111, and $T_{\rm eff}$ = 5460$\pm$100$\,K$ and $v \sin i$ = 13$\pm$2 km s$^{-1}$ for VSB 126.

The stability of the zero-point of the velocity system (e.g., Latham 1992) was monitored by taking exposures of the dusk and dawn sky, and applying small run-to-run corrections as described by Latham (1992). The final heliocentric RVs including these corrections are listed in Tables \ref{tbl:vsb111visiblerv} and \ref{tbl:vsb126visiblerv}. The RVs are plotted in Figures 5 and 6. 

The individual RVs for the stars based on the IR data were determined by using a two-dimensional cross correlation algorithm developed at Lowell Observatory following  Zucker \& Mazeh (1994). The algorithm calculates the correlation of the target spectrum against two templates chosen to best match the primary and secondary components. Approximate spectral types for the VSB 111 and VSB 126 primary
stars were known from visible light observations (Mathieu 1994).
To begin the process of estimating secondary star spectral types, we used the mass function from Mathieu (1994) to find the minimum secondary mass by assuming a primary mass of 1.0 M$_\odot$ for both VSB 111 and VSB 126. We estimated the secondary spectral type from the minimum secondary mass using model data presented in Luhman et al. (2003; Figure 5). We then used template spectra with the corresponding spectral type for each component taken from the suite presented in Prato et al. (2002). The two templates were shifted in RV, combined, and correlated with the observed spectra for each epoch, enabling the identification of component RVs. The flux ratio, $\alpha$, was determined by maximizing the cross-correlation coefficient for each epoch and taking the average. Subsequently, the average flux ratios, 0.39$\pm$0.05 for VSB 111 and 0.13$\pm$0.05 for VSB 126, were held fixed and the RVs were redetermined. The rotational velocity, $v \sin i$, was measured by using a set of templates rotationally broadened to a range of $v \sin i$ values (e.g., Mace et al. 2012). We selected the $v \sin i$ that yielded the maximum correlation coefficient. Tables \ref{tbl:vsb111irrv} and \ref{tbl:vsb126irrv} show the UT dates and measured IR RVs of both components of VSB 111 and VSB 126, respectively. These RVs are plotted in Figures 5 and 6. 

VSB 111 was found to best match a K0 template (BS 7368), with a $v \sin i$ of 30 km s$^{-1}$, and an M0 template (GL 763), with a $v \sin i$ of 15 km s$^{-1}$, for the primary and secondary components, respectively. 
VSB 126 was best fit with a K0 template (BS 7368), with a $v \sin i$ of 15 km s$^{-1}$, and an M4 template (GL 402), with a $v \sin i$ of 8 km s$^{-1}$, for the primary and secondary components, respectively.
These spectral types, the equivalent values of $T_{\rm eff}$ from Luhman et al. (2003), and
the $v \sin i$  values are all given in Table 9. Results from analysis of visible-light data are included for comparison. Conservative uncertainties of 5 km s$^{-1}$ are estimated for the IR $v \sin i$ values based on visual inspection of the spectra and comparison with standard star spectra convolved with different rotation kernels.

The uncertainties in the IR RVs for both VSB 111 and VSB 126 were initially set at
1.0 km s$^{-1}$ for the primary and 2.0 km s$^{-1}$ for the secondary (Prato et al. 2002).  
The uncertainties in the visible-light RVs for VSB 111
and VSB 126 were estimated by first using the internal
errors of the RVs as initial guesses. After finding the best orbital fit (\S 3.2) for the single-lined
data, we used the $\chi^2$ per degree of freedom as a guide to determine whether
these initial values were over or underestimated. The uncertainties were then
scaled by a multiplicative constant and the process repeated until we attained a 
$\chi^2$ per degree of freedom of $\sim$1.0. We then included the primary star IR RVs and
repeated the iterative process, multiplying the initial estimate of 1.0 km s$^{-1}$ by a constant
until we again attained a $\chi^2$ per degree of freedom of 1.0. Finally we included the
secondary star IR RVs and determined a full double-lined solution, again multiplying the
initial secondary uncertainties of 2.0 km s$^{-1}$ by a constant until the solution
converged on a $\chi^2$ per degree of freedom near unity. The final uncertainties
determined in visible light and in the IR are reported, respectively, in
Tables \ref{tbl:vsb111visiblerv} and \ref{tbl:vsb111irrv} for VSB 111 and in Tables \ref{tbl:vsb126visiblerv} and \ref{tbl:vsb126irrv} for VSB 126. 

\subsection{Orbital Parameters}

Because of the relatively large number of epochs, the visible light data for the primary star
dominate the orbital solutions for VSB 111 and VSB 126.  However,
combining our IR RVs for the primary and secondary with the visible RVs allows us to find
the double-lined solutions.  We used the Levenberg-Marquardt method with a standard
least-squares algorithm from Press et al. (1992). A search using a genetic algorithm (Charbonneau 1995) yielded an
initial single-lined binary solution based on the CfA spectra alone. The complete set of orbital parameters are listed 
in Table \ref{tbl:orbitalelements} for both spectroscopic binaries (SBs).  
 
Figures 5 and 6 show the combined orbital solutions for VSB 111 and VSB 126, respectively. We allowed the difference in the
center-of-mass velocity, $\gamma$, between the visible-light and IR RVs to be a free parameter in order to test the consistency
of the zero-point for the two data sets. For both systems we found an offset, $-$1.2 km s$^{-1}$ for VSB 111 and
$-$0.1 km s$^{-1}$ for VSB 126.  In the final, merged (visible$+$IR), double-lined solution we corrected the IR RVs for
these offsets before determining the final orbital parameters.  The offsets likely
arise as the result of small discrepancies in the RVs determined for the observed template
library spectra.

We calculated the O$-$C values for both SB2 systems and conducted a Scargle periodogram analysis on these residuals
to search for any underlying RV signals.  For both VSB 111 and VSB 126, the average O$-$C was close to zero.
No obvious periodicity was detected.

\subsection{Stellar Rotation Periods}

A Fourier-fitting routine searched for periodicities in the photometric data resulting from flux variations as the face
of the rotating star carries large spots across our line of sight. For VSB 111 we recover the previously published
rotation periods rather closely (Kearns \& Herbst 1998; Makidon et al. 2004);
our data indicate a period of 3.74$\pm$0.02 days (Table 9) with range
12.577 $< V <$ 12.650 mag.  Kearns \& Herbst published a period of 3.77 days and Makidon et al. published a
period of 3.75 days. The light curve of VSB 111 appears in Figure 3 and was fitted by a 
weighted, sinusodial curve producing an rms scatter of 0.006 mag. The data in Table \ref{tbl:vsb111photometry}
and Figure 3 show magnitudes binned in JD to minimize errors.
For VSB 126 the light curve was also fitted with a weighted sinusoid
to yield a rotation period of 5.71$\pm$0.07 days (Figure 4, Table 9). This is the first determined rotation period for VSB 126 in the literature.
The magnitude range on the fitted light curve is 13.499 $< V <$ 13.542 and the rms
scatter is 0.004 mag. Similarly, the data in Table \ref{tbl:vsb126photometry} and Figure 4 represent the binned results.
These rotation periods indicate the primaries are not pseudo-synchronized with the orbital
motions. Given the large primary to secondary star flux ratios in visible
light, which presumably prevented these systems from being identified as double-lined
binaries on the basis of the shorter wavelength data, we interpret these rotation periods as
pertaining to the primary star only.  

\subsection{Other Derived Parameters}

Luminosity was determined as
described in Prato et al. (2003, \S 3.3) using $T_{\rm eff}$, A$_V$ (determined from the
near-IR colors, as per Prato et al. 2003, to be zero for both VSB systems), r$_{K}$, the K-band excess, which was
zero for both our systems, a distance to NGC 2264 of 800 pc (Walker 1956), and the $J$-, $H$-,
and $K$-band values from 2MASS. A blackbody curve appropriate to the specific $T_{\rm eff}$ was produced, fit to
the $J$-, $H$-, and $K$-band magnitudes, and converted to absolute flux using d = 800 pc. We then integrated under this
curve to obtain the luminosity, L.  Errors for log(L/L$_\odot$) were determined from the uncertainties in distance and total
flux across the three bandpasses. To apportion the apparent magnitudes
between the primary and secondary stars, we applied the $H$-band flux ratio, $\alpha$, found in the
cross-correlation analysis, to the 2MASS magnitude.  Once the component $H$ magnitudes
were established, we used the $J-H$ and $H-K$ colors for
the relevant spectral types (Tokunaga 2000) to determine the $J$ and
$K$ component magnitudes. This yielded a similar result as applying the $H$-band flux
ratio from the cross-correlation to all three bandpasses, $J$, $H$, and $K$. The conservative errors adopted on the flux ratios
account for the largest source of error when determining the component apparent magnitudes. 
We estimated the primary star radii for both systems based on $T_{\rm eff}$ and luminosity, using
$L=4\pi\sigma R^{2}T_{\rm eff}^{4}$. The distance was the largest source of error in the luminosity calculations and the luminosity was the largest source of error in the radii calculations. Derived quantities for the components of VSB 111 and 126, such as L and R, are shown in Table \ref{tbl:physicalproperties}. We repeated this analysis for the primary stars in both systems based on the visible-light parameters as well (Table 9). 

With values for the primary star $v \sin i$, radius, and P$_{rot}$ we used the
relation $\sin i = v \sin i \times P_{rot} / (2\pi R)$ to determine the inclination of the
primary star rotation axes ($i_{stellar}$) with respect to the line of sight. We found $43^{+14}_{-12} \,^{\circ}$
for VSB 111 and $49^{+24}_{-23} \,^{\circ}$ for VSB 126 based on IR data and 45$\pm5 \,^{\circ}$ and $47^{+9}_{-8} \,^{\circ}$ based on the visible-light data. The largest source of error in the visible-light stellar inclination angle for VSB 111 stems from the uncertainty in radius and the largest source of error for VSB 126 stems from the rotation velocity, $\pm$2 km s$^{-1}$. The largest source of error in the IR-determined stellar inclination angles stems from the adopted error on the rotation velocity, $\pm$5 km s$^{-1}$, for both systems. Obtaining more precise values for the distance to VSB 111 and VSB 126 will help reduce the large errors in luminosity, radii, absolute magnitudes, and stellar inclination.

\section{Discussion}

\subsection{Masses and Ages}

To place the components of the systems on the H-R diagram and estimate the ages and masses of the stars, we tested four
sets of PMS evolutionary tracks, by Baraffe et al. (1998), Siess et al. (2000), Dotter et al. (2008), and Tognelli et al. (2011).  We used the values
of $T_{\rm eff}$ determined from the IR data, described above and given in Table 9, because the IR analysis successfully detected both SB components. 
For the Dotter et al. and Baraffe et al. tracks we plotted absolute $H$ magnitude to determine the stars' locations in the
H-R diagram.   For the Siess et al. and Tognelli et al. tracks we used log(L/L$_\odot$). For both components
in each system we used the distance of 800 pc to convert from apparent to absolute $H$ magnitude and to calculate
luminosities as described in \S 3.4 (Table 9).  Tables 11 and 12 list the results for
$M_{1}$ and $M_{2}$ as well as the corresponding mass ratios and approximate ages for VSB 111 and VSB 126, respectively. The
Dotter et al. tracks are plotted in Figure 7. 

The model results for mass are consistent with each other, which is not surprising given the large uncertainties involved in placing the
target components on the H-R diagram.  The visible light data yield independent values for $T_{\rm eff}$ for the primary stars in both systems
(\S 3.1) so it is possible to test our results using these numbers as well as the IR data.
For VSB 111, there is good agreement between the visible light and IR analysis for the primary star; substituting the value of $T_{\rm eff}$
from visible light spectroscopy does not change the H-R diagram results.  For VSB 126 the visible light $T_{\rm eff}$ is 5460 K,
over 200 K warmer than the IR value of 5248 K (Table 9); however, the only resulting change
from using this warmer $T_{\rm eff}$ in the H-R diagram analysis is to increase the age of the primary
slightly.  The VSB 126 primary star mass estimate remains unchanged as the tracks in this region of the H-R diagram are relatively horizontal.
A discrepancy of $>$2 $\sigma$ in the mass ratio is seen between all the evolutionary models and the mass ratio measured from the
orbital solutions for VSB 111; the models yield significantly lower mass ratios than the orbital solution. The models also
yield different ages for the individual components in both binaries.  The age determinations
for the secondary stars are consistently younger than the primary components, a systematic
bias that follows the trend identified in large population studies (e.g., Hillenbrand 1997).  Furthermore, the absolute age for the VSB 111 secondary,
$<$1 Myr for all models tested (Table 11), is unrealistically low.  However, given the very large uncertainties in the component ages,
these results are not significant; for the
Dotter et al. (2008) tracks we find age uncertainties of at least $\pm$1 Myr but typically of several Myr (Figure 7).

Torres et al. (2013) studied the four stars in the LkCa 3 system, a young hierarchical quadruple consisting of two short-period
spectroscopic binaries.  They found a lack of consistency between the dynamically determined orbital and physical
characteristics and the properties determined by comparing the
location of these stars in the H-R diagram with the models of Baraffe et al. (1998).  However, comparison of the Torres et al. observations
with the tracks of Dotter et al. (2008) showed good agreement, including between the measured
and track-derived mass ratios.  The $>$2 $\sigma$ inconsistency described above between the VSB 111 dynamically determined
mass ratio and that determined from comparison with the Dotter et al. tracks (Table 11) is thus puzzling. 
The spectral type, however, may provide a clue.  The primary star in VSB 111, K0, is
significantly warmer than those of the LkCa 3 primaries, which are between K7 and M2 (Torres et al.).  The Dotter et al.
tracks may thus provide consistent results between stars of lower masses but the higher mass primaries in the VSB systems
possibly span a range of mass tracks for which the results are less homogeneous.  This underscores the necessity of testing
tracks on a relatively broad sample of systems with a range of masses and mass ratios.

Far more stringent tests of evolutionary tracks would be possible with improvements in the
determination of $T_{\rm eff}$ and distance for the stellar components in VSB 111 and VSB 126.
Currently, our procedure is to use spectral type standards for the cross-correlation and to identify the young SB
components' spectral types from maximizing the correlation coefficient.  $T_{\rm eff}$ is then determined from compilations of
spectral type $-$ $T_{\rm eff}$ equivalence, such as those presented in Luhman (1999) and White et al. (1999).
To avoid this complication, Torres et al. (2013) adopted the $T_{\rm eff}$ values directly from those assigned by Rojas-Ayala et al. (2012) to 
several of the template stars used, or from the color/temperature calibrations of Boyajian et al. (2012).
In addition to the lack of $T_{\rm eff}$ values for all of our templates from Rojas-Ayala et al. and the lack of colors for the secondary stars in our
sample, it is not possible to determine $T_{\rm eff}$ for the primary stars in our sample in this way because they are too hot.  Thus
we have followed our standard procedure.  With a modest investment of observing time in the future, it would be possible
to obtain low-resolution ($\sim$2000) K-band spectroscopy and to follow the methodology of Rojas-Ayala et al. and assign
$T_{\rm eff}$ values directly to all low-mass ($T_{\rm eff}<4000$) stars in our library of template spectra.  An alternative
approach must be developed for earlier spectral type stars.

The determination of much improved distance estimates to the target SBs in NGC 2264 is also possible.
Combining the angular and physical scales of a system identified as both a visual binary and an SB2
yields an independent distance measurement.
For the 902 day period binary VSB 111, VLBA interferometry permits spatial mapping of the binary orbit
and will be possible if at least one of the components is a sufficiently strong radio source ($\ga$0.5 mJy), likely given that
VSB 111 has a large x-ray flux (Dahm et al. 2007).  Based on the orbital period and the total binary mass
determined from the Dartmouth tracks, Figure 7, and assuming a distance of 800 pc,
the average stellar separation is $\sim$3 mas, easily resolvable by the 0.8 mas FWHM VLBA beam at 8 GHz (4 cm).  For VSB 126,
GAIA is sufficiently sensitive (standard errors of 15$-$20 $\mu$as for the $V=13.4$ mag of VSB 126; de Bruijne 2012) to resolve the orbit:
based on the total mass estimated from the Dartmouth tracks, the period, and d$=$800 pc, the average stellar separation of VSB 126 is 165 $\mu$as.
GAIA is scheduled for an October, 2013, launch.

\subsection{Rotational Velocities}

Based on the IR analysis, VSB 111 has a primary rotational velocity of 30 km s$^{-1}$ with a secondary rotational velocity of 15 km s$^{-1}$ and VSB 126 has a primary rotational velocity of 15 km s$^{-1}$ and a secondary rotational velocity of 8 km s$^{-1}$ (Table 9). These are consistent to within 1$\sigma$ of the primary star $v \sin i$ values found in the visible light analysis (section 3.1). We searched the literature for other $v \sin i$ measurements in PMS SB2s (e.g., Ru\'{i}z-Rodr\'{i}guez et al. 2013; Landin et al. 2009; Reipurth et al. 2002; Covino et al. 2001) and found $\sim$35 systems for which the $v \sin i$ values of both SB2 components have been measured. Half of these systems are isolated and half are in young clusters. We examined the ratio $v_1 \sin i$ to $v_2 \sin i$ as a function of mass ratio, orbital period, and eccentricity and found that eight systems
have a primary star $v \sin i$ value about twice as great as the secondary $v \sin i$ value, including VSB 111 and 126 (Table \ref{tbl:vsini}). 
For the remainder of the systems examined the ratio was close to unity.
There were no obvious correlations found when comparing this ratio to $e$, $q$, and orbital period.
This result might suggest that rotational evolution between a $v \sin i$ ratio of  $\approx$ 2
and a ratio of $\approx$ 1 takes place rapidly, or it may simply be the outcome of observational bias and/or imprecise assignation of $v \sin i$ values.
A comparable sample of $v \sin i$ values for main sequence SB2s, from Goldberg et al. (2002), showed no preferential distribution in the primary to
secondary star $v \sin i$ ratio.  Given the large uncertainties associated with most of the young star $v \sin i$ values from the literature, as well as with
our IR results for VSB 111 and VSB 126, i.e. $\pm$5 km s$^{-1}$, the odd distribution of $v \sin i$ ratios in young SBs is unlikely to be significant.

\subsection{Additional Orbital Properties}

We used the model tracks shown in Figure 7 and estimated the mass and associated
uncertainties for each star in each binary (Tables 11 and 12).  Given
the values for $M_1 \sin ^{3} i$ and $M_2 \sin ^{3} i$ determined in the orbital fit
for each SB2 (Table 10), we combined these to estimate $i_{orb}$ as illustrated in Figures 8 and 9 for VSB 111 and 126, respectively
(Prato et al. 2001; Mace et al. 2012). From these plots we find that for VSB 111 $i_{orb} = 48\,^{\circ}$ and
for VSB 126 $i_{orb}$ is between $36\,^{\circ}$ and $39\,^{\circ}$.

In \S 3.4 we estimated the inclinations of the VSB 111 and VSB 126 primary star rotation
axes to be 45$\pm5\,^{\circ}$ and $47^{+9}_{-8}\,^{\circ}$, respectively, on the basis of the visible-light data analysis.
Unfortunately, without additional information about the visual orbits of these systems, i.e. knowledge of
the nodal angles, it is not possible to determine whether there is agreement between the stellar equatorial planes and
the orbital planes.  However, with facilities such as the VLBA and the GAIA mission, due for launch late in 2013,
it will be possible to map out these orbits
and to determine the relative orientation of the orbital plane with respect to the stellar
equatorial plane.  These relative orientations are important for understanding the dynamics of
star-orbit interactions, which may in turn shape the planet formation and evolution environment.
By directly and accurately measuring the orbital inclination, it will also be possible to derive
the model-independent absolute component masses directly.

\subsection{NGC 2264}

The locations of VSB 111 and 126 are separated by $\sim$0.02 pc in right ascension and $\sim$1.9 pc in declination, assuming an 800 pc distance to NGC 2264.
VSB 111, with $\gamma$ = 25.31 km s$^{-1}$ (Table \ref{tbl:orbitalelements}), lies very close to the core of a $^{13}$CO cloud, as determined from visual inspection of Figure 1 from F\H{u}r\'{e}sz et al. (2006).  VSB 126, with $\gamma$ = 17.16 km s$^{-1}$ (Table \ref{tbl:orbitalelements}), does not appear to be associated with any cloud cores.
If VSB 111 is actually associated with this cloud core, this location is
consistent with the age of VSB 111 being younger than that of VSB 126, as suggested by Figure 7.

Both $\gamma$ velocities for VSB 111 and VSB 126 fall within the cluster distribution range for membership: F\H{u}r\'{e}sz et al. (2006) found that NGC 2264 is hierarchically structured with a north-south RV gradient, which increases with increasing declination. Stars in the northern half of the cluster have an overall higher RV than the stars in the southern half. The mean RV for stars with declination near +9$\,^{\circ}$ $50\arcmin$ \ is $\sim$27 km s$^{-1}$ and the mean RV for stars with declination near +9$\,^{\circ}$ $37\arcmin$ \ is $\sim$16 km s$^{-1}$. Stars that are in the same region of NGC 2264 as VSB 111 and VSB 126 thus have RVs that are consistent with these binaries' $\gamma$ velocities.

Interstellar reddening toward NGC 2264 is fortuitously negligible.  This not only ensures that the targets are as bright as possible given their relatively large distances but also facilitates the measurement of luminosity, which may be hindered by the accurate determination
of A$_V$. The largest value of reddening was determined by Rebull et al. (2002) to be E(B-V)=0.15$\pm$0.03 for a distance of 760 pc. The lowest value of reddening was determined by Feldbrugge \& van Genderen (1991) to be E(B-V)=0.04 at a distance of 700 pc. There is a dark cloud of dense dust that lies directly behind NGC 2264, reducing the possibility of background stars contaminating the field.  The low value for extinction found in the literature agree with the values we calculated of A$_V=0$ for both systems (\S 3.4).

\section{Summary}

High-resolution, IR spectra were obtained for the young SBs VSB 111 and VSB 126 in NGC 2264 and combined
with visible light spectra. One- and two-dimensional cross-correlation was used to obtain the RVs and orbital parameters for these systems. 
VSB 111 has a period of $\sim$902 days, an eccentricity of 0.79$\pm$0.01, and a mass ratio of 0.52$\pm$0.05 while VSB 126 has a period of $\sim$12.9 days,
an eccentricity of 0.18$\pm$0.02, and a mass ratio of 0.29$\pm$0.02. The eccentricity of VSB 111 is the
second highest known for a PMS SB2 measured to date (Mace et al. 2009).
We determined the stellar rotation periods for the primary components of both systems, $\sim$4 and $\sim$6 days
for VSB 111 and VSB 126, respectively.  

Our analysis indirectly provided additional physical properties for both
stars in each SB2, such as $T_{\rm eff}$, binary flux ratio, luminosity,
and $v \sin i$.  These results in turn allowed us to place both stars in each binary on an H-R diagram
for comparison with model calculations of PMS evolution, effectively testing the evolutionary tracks against our
dynamical results.  By using the model-determined masses, we estimated the binaries' orbital inclinations;
we also found the inclination angles of the primary
stars' equatorial planes.  With observations of the angularly resolved orbits of these NGC 2264 members, through VLBI
and GAIA, the dynamical determination of the orbital inclination and orientation (i.e., the nodal angle) will be possible.
These measurements in turn yield the absolute component masses, distance to the system, and the relative inclinations 
of the primary stars with respect to the orbital planes.

Determining the spectroscopic orbital solutions for these two young SB2s inches us closer to our immediate goal to significantly
increase the number of such systems in order to better understand and improve models of binary star formation, and to help
anchor theoretical evolutionary tracks with solid dynamical data.

\vspace{12 pt}

We are grateful to G. Mace, M. Simon, K. Covey, and J. Patience for useful discussions and comments throughout the progress of
this project. We thank Joel Aycock, Gary Puniwai, Gabrelle Saurage, and Cynthia Wilburn for their
superb telescope support and Randy Campbell, Al Conrad, Jim Lyke, Barbara Schaefer, and
Greg Wirth for their dedicated technical and logistical support.  We are grateful to P. Berlind, J. Caruso, R. J. Davis,
L. Hartmann, E. Horine, A. Milone, J. Peters, J. Stauffer, R. P. Stefanik, and S. Tokarz for help in
obtaining the visible-light spectra of VSB 111 and VSB 126. This work was supported in part by
the NSF grant AST-1009136 to L.P.; G.T. acknowledges partial support from NSF grant AST-1007992.
Data presented herein were obtained at the W. M. Keck Observatory from telescope time allocated to
the National Aeronautics and Space Administration through the agencyÕs scientific partnership with
the California Institute of Technology and the University of California. Keck telescope time was also
granted by NOAO, through the Telescope System Instrumentation Program (TSIP). TSIP is funded by
NSF. The Observatory was made possible by the generous financial support of the W. M. Keck Foundation.
This work made use of the SIMBAD database, the Vizier database, the NASA
Astrophysics Data System, and the data products from the Two Micron All Sky Survey, which is a joint
project of the University of Massachusetts and the Infrared Processing and Analysis Center/California
Institute of Technology, funded by the National Aeronautics and Space Administration and the NSF.
This publication also made use of data products from the Wide-field Infrared Survey Explorer, which is a
joint project of the University of California, Los Angeles, and the Jet Propulsion Laboratory/California
Institute of Technology, funded by the National Aeronautics and Space Administration. We recognize
and acknowledge the significant cultural role that the summit of Mauna Kea plays within the indigenous
Hawaiian community and are grateful for the opportunity to conduct observations from this special mountain.

\newpage 

\begin{deluxetable}{ccc}
\tablewidth{0pt}
\tablecaption{Target Properties
 \label{tbl:targetproperties}}
\tablehead{
\colhead{Property}  & \colhead{VSB 111} & \colhead{VSB 126}}
\startdata
R.A. (J2000) & 06 41 04.41 & 06 41 07.778 \\
Dec. (J2000) & +09 51 50.1 & +09 44 03.00 \\
$U$ (mag)\tablenotemark{a} & 13.677$\pm$0.004 & 14.391$\pm$0.031 \\
$B$ (mag)\tablenotemark{b} & 13.60 & 14.17 \\
$V$ (mag)\tablenotemark{b} & 12.68 & 13.39 \\
$J$ (mag) & 10.737$\pm$0.023 & 11.887$\pm$0.022 \\
$H$ (mag)  & 10.300$\pm$0.024 & 11.513$\pm$0.026 \\
$K$ (mag)  & 10.145$\pm$0.027 & 11.383$\pm$0.025 \\
H$\alpha$ EW (\AA)\tablenotemark{c}  & -0.3 &-1.9 \\ 
\enddata

\tablenotetext{a}{Sung, Bessell, \& Lee (1997)}
\tablenotetext{b}{Sagar \& Joshi (1983)}
\tablenotetext{c}{Mathieu (1994)}

\end{deluxetable}

\newpage

\begin{deluxetable}{ccc}
\tablewidth{0pt}
\tablecaption{VSB 111 Visible-Light Radial Velocities
	\label{tbl:vsb111visiblerv}}
\tablehead{
\colhead{HJD} & \colhead{$v$$_{1}$$\pm$$\sigma$ (km s$^{-1}$)}  & \colhead{Phase}}

\startdata
2445714.8925 & 21.3$\pm$0.9 & 0.1975 \\
2445747.6330 & 19.9$\pm$1.8 & 0.2338 \\
2446066.9770 & 26.0$\pm$1.4 & 0.5878 \\
2446076.8868 & 26.1$\pm$1.0 & 0.5988 \\
2446420.9011 & 32.9$\pm$1.7 & 0.9802 \\
2446489.7238 & 14.5$\pm$1.2 & 0.0565 \\
2446776.7760 & 25.4$\pm$1.5 & 0.3747 \\
2446802.9369 & 26.5$\pm$1.2 & 0.4037 \\
2446814.9734 & 23.1$\pm$1.2 & 0.4170 \\
2446873.6357 & 25.7$\pm$1.2 & 0.4821 \\
2446873.6538 & 25.5$\pm$2.4 & 0.4821 \\
2447074.0159 & 30.1$\pm$1.5 & 0.7042 \\
2447076.9919 & 29.5$\pm$0.8 & 0.7075 \\
2447079.9911 & 29.9$\pm$1.6 & 0.7108 \\
2447127.9563 & 32.2$\pm$1.3 & 0.7640 \\
2447128.8350 & 29.9$\pm$1.6 & 0.7650 \\
2447138.8504 & 29.9$\pm$1.0 & 0.7761 \\
2447157.9513 & 29.7$\pm$1.2 & 0.7972 \\
2447169.9786 & 31.1$\pm$1.2 & 0.8106 \\
2447192.6830 & 31.4$\pm$1.6 & 0.8357 \\
2447220.7105 & 31.4$\pm$1.3 & 0.8668 \\
2447250.7112 & 30.2$\pm$1.4 & 0.9001 \\
2447428.9314 & 15.5$\pm$1.1 & 0.0976 \\
2447489.8353 & 20.6$\pm$1.4 & 0.1651 \\
2447493.8657 & 20.8$\pm$1.0 & 0.1696 \\
2447523.9220 & 24.5$\pm$1.1 & 0.2029 \\
2447549.9107 & 23.8$\pm$0.8 & 0.2317 \\
2447573.7981 & 22.0$\pm$1.4 & 0.2582 \\
2447580.7045 & 23.5$\pm$1.0 & 0.2659 \\
2447641.6147 & 23.0$\pm$1.8 & 0.3334 \\
2447642.6205 & 22.8$\pm$1.5 & 0.3345 \\
2447791.9773 & 25.8$\pm$1.1 & 0.5001 \\
2447811.9797 & 26.3$\pm$1.6 & 0.5223 \\
2447812.9739 & 25.2$\pm$1.1 & 0.5234 \\
2447845.9494 & 28.6$\pm$1.0 & 0.5599 \\
2447898.8962 & 26.9$\pm$1.1 & 0.6186 \\
2447906.7812 & 27.8$\pm$1.4 & 0.6274 \\
2447958.7219 & 28.9$\pm$1.1 & 0.6849 \\
2448168.9877 & 33.5$\pm$1.1 & 0.9180 \\
2448200.9677 & 33.6$\pm$1.5 & 0.9535 \\
2448256.7928 & 9.3$\pm$1.2 & 0.0154 \\
2448261.6991 & 10.0$\pm$1.9 & 0.0208 \\
2448281.7687 & 13.8$\pm$1.5 & 0.0208 \\
2448284.7245 & 10.9$\pm$1.4 & 0.0463 \\
2448288.7069 & 13.9$\pm$0.8 & 0.0507 \\
2448340.6732 & 18.5$\pm$1.4 & 0.1084 \\
2448346.6417 & 18.2$\pm$0.9 & 0.1150 \\
2448347.6792 & 20.0$\pm$1.4 & 0.1161 \\
2448367.6272 & 18.7$\pm$2.3 & 0.1382 \\
2448370.6468 & 17.3$\pm$2.1 & 0.1416 \\
2448669.7119 & 26.6$\pm$1.0 & 0.4731 \\
2448671.8058 & 26.4$\pm$1.2 & 0.4754 \\
2448910.9782 & 30.8$\pm$1.0 & 0.7406 \\
2449647.9617 & 26.8$\pm$1.1 & 0.5576 \\
2449701.9633 & 25.2$\pm$1.3 & 0.8174 \\
2450001.9827 & 32.7$\pm$0.8 & 0.9500 \\
2450026.8712 & 31.7$\pm$1.9 & 0.9776 \\
2450029.9581 & 29.2$\pm$0.9 & 0.9810 \\
2450034.9645 & 27.5$\pm$1.0 & 0.9866 \\
2450037.0311 & 25.4$\pm$1.3 & 0.9889 \\
2450038.8478 & 24.8$\pm$1.6 & 0.9909 \\
2450051.9235 & 11.3$\pm$1.0 & 0.0054 \\
2450059.8782 & 10.3$\pm$1.0 & 0.0142 \\
2450064.8144 & 9.0$\pm$1.5 & 0.0197 \\
2450080.7933 & 10.5$\pm$3.0 & 0.0374 \\
2450081.8739 & 13.2$\pm$1.6 & 0.0386 \\
2450084.8209 & 10.5$\pm$0.9 & 0.0418 \\
2450091.8166 & 13.5$\pm$1.3 & 0.0496 \\
\enddata
\end{deluxetable}

\newpage

\begin{deluxetable}{ccc}
\tablewidth{0pt}
\tablecaption{VSB 126 Visible-Light Radial Velocities
	\label{tbl:vsb126visiblerv}}
\tablehead{
\colhead{HJD} & \colhead{$v$$_{1}$$\pm$$\sigma$ (km s$^{-1}$) } & \colhead{Phase}}

\startdata
2445783.6412 & 9.0$\pm$1.0 & 0.1224 \\
2446077.7737 & 5.2$\pm$0.9 & 0.8804  \\
2446775.9909 & 3.1$\pm$0.7 & 0.9038  \\
2447075.9919 & 7.2$\pm$0.9 & 0.1159  \\
2447078.9952 & 25.9$\pm$1.1 & 0.3483  \\
2447127.9675 & 11.5$\pm$1.0 & 0.1374 \\
2447128.8485 & 19.2$\pm$1.0 & 0.2056 \\
2447138.8614 & 0.7$\pm$1.0 & 0.9803 \\
2447157.9395 & 30.1$\pm$0.6 & 0.4564 \\
2447158.9747 & 27.9$\pm$0.8 & 0.5365 \\
2447159.9427 & 26.8$\pm$0.8 & 0.6114 \\
2447169.9634 & 28.5$\pm$1.3 & 0.3868 \\
2447199.7030 & 22.8$\pm$1.3 & 0.6878 \\
2447200.7497 & 14.6$\pm$1.1 & 0.7688 \\
2447220.6977 & 26.8$\pm$0.8 & 0.3122 \\
2447222.7006 & 28.5$\pm$1.5 & 0.4672 \\
2447224.6626 & 22.6$\pm$2.1 & 0.6190 \\
2447229.6414 & -0.1$\pm$1.0 & 0.0043 \\
2447230.6267 & 5.0$\pm$0.9 & 0.0805 \\
2447250.7040 & 24.5$\pm$0.7 & 0.6339 \\
2447427.9928 & 26.4$\pm$0.9 & 0.3514 \\
2447428.9520 & 29.2$\pm$1.0 & 0.4256 \\
2447429.9625 & 28.9$\pm$0.9 & 0.5038 \\
2447489.8436 & 12.6$\pm$0.8 & 0.1370 \\
2447493.8731 & 29.0$\pm$0.8 & 0.4487 \\
2447524.9283 & 6.5$\pm$0.9 & 0.8516 \\
2447549.9192 & 14.4$\pm$0.6 & 0.7852 \\
2447845.9554 & 21.8$\pm$0.9 & 0.6905 \\
2448669.7216 & 30.3$\pm$1.3 & 0.4280 \\
\enddata

\end{deluxetable}

\newpage

\begin{deluxetable}{ccccc}
\tablewidth{0pt}
\tablecaption{VSB 111 IR Radial Velocities
	\label{tbl:vsb111irrv}}
\tablehead{
\colhead{UT Date of Observations} & \colhead{HJD} & \colhead{$v$$_{1}$$\pm$$\sigma$ (km s$^{-1}$) } & \colhead{$v$$_{2}$$\pm$$\sigma$ (km s$^{-1}$) } & \colhead{Phase}}

\startdata
2001 Dec 31 & 2452274.94279 & 25.6$\pm$1.0& 24.4$\pm$2.0 & 0.4697 \\
2002 Feb 6 & 2452311.86271 & 27.1$\pm$1.0& 23.5$\pm$2.0 & 0.5106 \\
2002 Dec 14 & 2452623.04253 & 32.7$\pm$1.0 & 14.4$\pm$2.0 & 0.8556 \\
2004 Jan 28 & 2453032.84263 & 23.6$\pm$1.0& 28.5$\pm$2.0 & 0.3099 \\
2004 Dec 26 & 2453366.04274 & 27.7$\pm$1.0& 20.5$\pm$2.0 & 0.6793 \\
2010 Dec 10 & 2455543.11876 & 15.7$\pm$1.0 & 43.3$\pm$2.0 & 0.0927 \\
2012 Jan 11 & 2455937.93446 & 26.3$\pm$1.0 & 23.4$\pm$2.0 & 0.5304 \\
\enddata

\end{deluxetable}

\newpage

\begin{deluxetable}{ccccc}
\tablewidth{0pt}
\tablecaption{VSB 126 IR Radial Velocities
	\label{tbl:vsb126irrv}}
\tablehead{
\colhead{UT Date of Observations} & \colhead{HJD} & \colhead{$v$$_{1}$$\pm$$\sigma$ (km s$^{-1}$)} & \colhead{$v$$_{2}$$\pm$$\sigma$ (km s$^{-1}$)} & \colhead{Phase}} 

\startdata
2002 Jan 1 & 2452275.91504 & 29.1$\pm$1.0 & -25.3$\pm$5.0 & 0.4511 \\
2002 Feb 5 & 2452310.90655 & 13.5$\pm$1.0 & 18.4$\pm$5.0 & 0.1585 \\
2002 Dec 14 & 2452623.07869 & 24.4$\pm$1.0 & -15.7$\pm$5.0 & 0.3123 \\
2002 Dec 22 & 2452631.03900 & 1.3$\pm$1.0 & 71.7$\pm$5.0 & 0.9282 \\
2004 Jan 27 & 2453031.95005 & 1.3$\pm$1.0 & 76.5$\pm$5.0 & 0.9481 \\
2004 Dec 25 & 2453365.06614 & 20.9$\pm$1.0 & 11.3$\pm$5.0 & 0.7224
\enddata

\end{deluxetable}

\newpage

\begin{deluxetable}{cccccccc}
\tablewidth{0pt}
\tablecaption{Photometric Comparison Stars
	\label{tbl:compstars}}
\tablehead{
\colhead{ID} & \colhead{Walker} & \colhead{RA (J2000)} & \colhead{Dec (J2000)} & \colhead{$V$(mag)} & \colhead{$B$ - $V$} &\colhead{aliases}}

\startdata
1 & NGC 2264 086 & 6 40 37.49 & +09 54 57.8 & 11.73 & 0.63 & NSV 17013 \\
2 & NGC 2264 216 & 6 41 31.50 & +09 54 54.8 & 11.90 & 0.77 & NSV 3156 \\
3 & NGC 2264 190 & 6 41 31.81 & +09 55 43.9 & 12.34 & 0.62 & NSV 17104 \\
4 & NGC 2264 125 & 6 40 56.97 & +09 48 40.7 & 12.32 & 0.60 & NSV 17059 \\
5 & NGC 2264 151 & 6 41 02.96 & +09 47 54.3 & 12.57 & 0.49 & NSV 3117 
\enddata

\end{deluxetable}

\newpage

\begin{deluxetable}{ccc}
\tablewidth{0pt}
\tablecaption{VSB 111 Photometry
	\label{tbl:vsb111photometry}}
\tablehead{
\colhead{BJD} & {$V$ mag $\pm$$\sigma$} & {Phase}}

\startdata
2456204.87406    &  12.577$\pm$0.005 & 0.033 \\
2456204.89336    &  12.581$\pm$0.005 & 0.038 \\
2456204.93308    &  12.590$\pm$0.004 & 0.049 \\
2456204.95158    &  12.587$\pm$0.004 & 0.054 \\
2456204.96819    &  12.590$\pm$0.004 & 0.058 \\
2456204.98598    &  12.590$\pm$0.004 & 0.631 \\
2456205.01121    &  12.590$\pm$0.004 & 0.070 \\
2456205.01694    &  12.588$\pm$0.004 & 0.071 \\
2456205.87830    &  12.642$\pm$0.004 & 0.302 \\
2456205.93331    &  12.645$\pm$0.004 & 0.316 \\
2456205.94994    &  12.644$\pm$0.004 & 0.321 \\
2456205.97125    &  12.646$\pm$0.004 & 0.327 \\
2456205.98655    &  12.646$\pm$0.004 & 0.331 \\
2456206.01396    &  12.650$\pm$0.004 & 0.338 \\
2456206.01763    &  12.648$\pm$0.004 & 0.339 \\
2456207.85301    &  12.619$\pm$0.005 & 0.830 \\
2456207.92283    &  12.617$\pm$0.005 & 0.848 \\
2456207.95936    &  12.613$\pm$0.005 & 0.858 \\
2456208.01980    &  12.607$\pm$0.004 & 0.874 \\
2456208.88059    &  12.604$\pm$0.004 & 0.104 \\
2456208.94489    &  12.610$\pm$0.004 & 0.122 \\
2456208.97615    &  12.610$\pm$0.004 & 0.130 \\
2456209.87799    &  12.640$\pm$0.004 & 0.371 \\
2456209.94233    &  12.639$\pm$0.004 & 0.388 \\
2456209.97354    &  12.636$\pm$0.007 & 0.397 \\
2456210.87131    &  12.642$\pm$0.004 & 0.637 \\
2456210.93555    &  12.639$\pm$0.004 & 0.654 \\
2456210.96680    &  12.644$\pm$0.004 & 0.662 \\
2456211.01759    &  12.643$\pm$0.004 & 0.676 \\
2456211.87116    &  12.601$\pm$0.004 & 0.904 \\
2456211.93582    &  12.602$\pm$0.004 & 0.921 \\
2456211.96794    &  12.602$\pm$0.004 & 0.930 \\
2456212.01888    &  12.600$\pm$0.003 & 0.944 
\enddata

\end{deluxetable}

\newpage

\begin{deluxetable}{ccc}
\tablewidth{0pt}
\tablecaption{VSB 126 Photometry
	\label{tbl:vsb126photometry}}
\tablehead{
\colhead{BJD} & \colhead{$V$ mag $\pm$$\sigma$} & \colhead{Phase}}
\startdata

2456204.87406 & 13.505$\pm$0.008 & 0.022 \\     
2456204.89336 & 13.506$\pm$0.008 & 0.025 \\     
2456204.93308 & 13.504$\pm$0.006 & 0.032 \\     
2456204.95158 & 13.503$\pm$0.006 & 0.035 \\     
2456204.96819 & 13.508$\pm$0.006 & 0.041 \\     
2456204.98598 & 13.502$\pm$0.005 & 0.046 \\     
2456205.01121 & 13.503$\pm$0.005 & 0.047 \\     
2456205.01694 & 13.504$\pm$0.006 & 0.198 \\     
2456205.87830 & 13.524$\pm$0.006 & 0.207 \\    
2456205.93331 & 13.520$\pm$0.006 & 0.210 \\     
2456205.94994 & 13.523$\pm$0.006 & 0.214 \\      
2456205.97125 & 13.515$\pm$0.005 & 0.217 \\     
2456205.98655 & 13.521$\pm$0.005 & 0.221 \\     
2456206.01396 & 13.520$\pm$0.005 & 0.222 \\     
2456206.01763 & 13.518$\pm$0.006 & 0.543 \\     
2456207.85301 & 13.536$\pm$0.009 & 0.556 \\     
2456207.92283 & 13.540$\pm$0.008 & 0.572 \\     
2456207.95936 & 13.542$\pm$0.008 & 0.723 \\     
2456208.01980 & 13.541$\pm$0.006 & 0.735 \\     
2456208.88059 & 13.527$\pm$0.006 & 0.740 \\      
2456208.94489 & 13.527$\pm$0.005 & 0.898 \\     
2456208.97615 & 13.518$\pm$0.005 & 0.909 \\     
2456209.87799 & 13.511$\pm$0.005 & 0.915 \\     
2456209.94233 & 13.506$\pm$0.006 & 0.072 \\     
2456209.97354 & 13.519$\pm$0.011 & 0.083 \\     
2456210.87131 & 13.503$\pm$0.005 & 0.098 \\     
2456210.93555 & 13.499$\pm$0.005 & 0.247 \\     
2456210.96680 & 13.505$\pm$0.005 & 0.258 \\     
2456211.01759 & 13.502$\pm$0.005 & 0.264  \\     
2456211.87116 & 13.520$\pm$0.005 & 0.273 \\     

\enddata

\end{deluxetable}

\newpage

\begin{deluxetable}{ccccc}
\tablewidth{0pt}
\tablecaption{Physical Properties of VSB 111 and VSB 126
	\label{tbl:physicalproperties}}
\tablehead{
\colhead{$ $} & \colhead{IR} & \colhead{IR} & \colhead{Visible} & \colhead{Visible} \\
\colhead{Property} & \colhead{VSB 111} & \colhead{VSB 126} & \colhead{VSB 111} & \colhead{VSB 126}}
\startdata
Primary Rotation Period (days) & $-$ & $-$ & 3.74$\pm$0.02 &  5.71$\pm$0.07\\
Primary Spectral Type & K0$\pm$1  & K0$\pm$1 & $-$ & $-$  \\
Secondary Spectral Type & M0$\pm$1 & M4$\pm$1 & $-$ & $-$ \\
Primary $T_{\rm eff}$ (K) & 5248$\pm$194 & 5248$\pm$194 & 5300$\pm$100 & 5460$\pm$100 \\
Secondary $T_{\rm eff}$ (K) & 3846$\pm$192 & 3258$\pm$180 & $-$ & $-$ \\
Primary $v \sin i$ (km s$^{-1}$) & 30$\pm$5 & 15$\pm$5 & 31$\pm$2 & 13$\pm$2 \\
Secondary $v \sin i$ (km s$^{-1}$) & 15$\pm$5 & 8$\pm$5 & $-$ & $-$ \\
log(L$_{1}$/L$_\odot$) & $0.87^{+0.16}_{-0.17}$ & 0.47$\pm0.17$ & 0.87$\pm$0.07 & $0.51^{+0.07}_{-0.08}$ \\
log(L$_{2}$/L$_\odot$) & $0.22^{+0.28}_{-0.19}$ & $-0.72^{+0.26}_{-0.34}$ & $-$ & $-$ \\
R$_{1}$ (R$_\odot$) & $3.29^{+0.88}_{-0.73}$ & $2.09^{+0.57}_{-0.47}$ & $3.25^{+0.29}_{-0.28}$ & $2.00^{+0.19}_{-0.17}$
\enddata

\end{deluxetable}

\newpage

\normalsize
\begin{deluxetable}{ccc}
\tablewidth{0pt}
\tablecaption{Orbital Elements and Properties for VSB 111 and VSB 126
	\label{tbl:orbitalelements}}
\tablehead{
\colhead{Element/Property} & \colhead{VSB 111} & \colhead{VSB 126}} 

\startdata
P (days) & 902.1$\pm$0.9 & 12.9244$\pm$0.0002  \\
$\gamma$ (km s$^{-1}$) & 25.31$\pm$0.16 & 17.16$\pm$0.17  \\
K$_{1}$ (km s$^{-1}$) & 12.00$\pm$0.32 & 14.74$\pm$0.24  \\
K$_{2}$ (km s$^{-1}$) & 23.1$\pm$2.2 & 51.3$\pm$2.8  \\
$e$ & 0.788$\pm$0.008 & 0.18$\pm$0.02  \\
$\omega$ (deg) & 115.6$\pm$2.2 & 190.1$\pm$5.3  \\
T +2,444,000(HJD) & 634.6$\pm$5.3 & 605.94$\pm$0.17  \\
$M_1 \sin ^{3} i$ (M$_\odot$) & 0.618$\pm$0.135 & 0.285$\pm$0.040  \\
$M_2 \sin ^{3} i$ (M$_\odot$) & 0.321$\pm$0.042 & 0.082$\pm$0.007  \\
q$=$$M_{2}$/$M_{1}$ & 0.52$\pm$0.05 & 0.29$\pm$0.02  \\
a$_{1}$sin $i$ (Gm) & 91.6$\pm$2.0 & 2.58$\pm$0.04  \\
a$_{2}$sin $i$ (Gm) & 176.1$\pm$16.3 & 8.96$\pm$0.49  \\

\enddata

\end{deluxetable}

\newpage

\begin{deluxetable}{cccccc}
\tablewidth{0pt}
\tablecaption{Evolutionary Track Comparisons for VSB 111 ($q_{obs}$ = 0.52$\pm$0.05)
	\label{tbl:11}}
\center
\tablehead{
\colhead{Tracks} & \colhead{M$_{1}$ (M$_\odot$)} & \colhead{Age (Myr)} & \colhead{M$_{2}$ (M$_\odot$)} & \colhead{Age (Myr)} & \colhead{$q\pm1\sigma$}}
\startdata
Baraffe et al. (1998)\tablenotemark{a}& - & - & 0.63$\pm$0.10 & $<$ 1 & -  \\
Tognelli et al. (2011) & 2.3$\pm$0.3 & 2$\pm$1 & 0.45$\pm$0.20 & 0.5$\pm$1.0 & 0.2$\pm$0.1  \\
Siess et al. (2000) & 2.1$\pm$0.3 & 3$\pm$2 & 0.41$\pm$0.10 & 0.3$\pm$0.5 & 0.2$\pm$0.1  \\
Dotter et al. (2008) & 2.2$\pm$0.3 & 2$\pm$1 & 0.47$\pm$0.15 & $<$1 & 0.2$\pm$0.1 \\
\enddata
\tablenotetext{a}{Baraffe et al. (1998) tracks ($\alpha$ = 1.9) do not go above 1.4M$_\odot$.}

\end{deluxetable}

\newpage
\tablewidth{0pt}
\begin{deluxetable}{cccccc}
\tablecaption{Evolutionary Track Comparisons for VSB 126 ($q_{obs}$ = 0.29$\pm$0.02)
	\label{tbl:12}}
\centering
\tablehead{
\colhead{Tracks} & \colhead{M$_{1}$ (M$_\odot$)} & \colhead{Age (Myr)} & \colhead{M$_{2}$ (M$_\odot$)} & \colhead{Age (Myr)} & \colhead{$q\pm1\sigma$}}
\startdata
Baraffe et al. (1998)\tablenotemark{a}  & - & - & 0.28$\pm$0.14 & 2$\pm$5 & - \\
Tognelli et al. (2011)  & 1.7$\pm$0.3 & 4$\pm$2 & 0.25$\pm$0.12 & 2$\pm$3 & 0.2$\pm$0.1 \\
Siess et al. (2000)  & 1.6$\pm$0.3 & 5$\pm$3 & 0.17$\pm$0.07 & 1$\pm$2 & 0.1$\pm$0.1 \\
Dotter et al. (2008) & 1.6$\pm$0.4 & 5$\pm$4 & 0.24$\pm$0.11 & 2$\pm$6 &0.2$\pm$0.1 \\
\enddata
\tablenotetext{a}{Baraffe et al. (1998) tracks ($\alpha$ = 1.9) do not go above 1.4M$_\odot$.}

\end{deluxetable}

\newpage

\begin{deluxetable}{ccccccc}
\tablewidth{0pt}
\tablecaption{Properties of PMS SB2s with $v_{1} \sin i$ twice as large as $v_{2} \sin i$
	\label{tbl:vsini}}
\tablehead{
\colhead{System} & \colhead{$v{_1} \sin i$ (km s$^{-1}$)} & \colhead{$v{_2} \sin i$ (km s$^{-1}$)} & \colhead{$e$} & \colhead{$q$} & \colhead{Period (days)}}
\startdata

VSB 111\tablenotemark{a} & 30$\pm$5 & 15$\pm$5 & 0.79 & 0.52 & 902.1 \\
VSB 126\tablenotemark{a} & 15$\pm$5 & 8$\pm$5 & 0.18 & 0.29 & 12.92 \\
RX J0539.9$+$0956\tablenotemark{b} & 80$\pm$10 & 40$\pm$5 & 0.27 & 0.7 & 1118.3 \\
RX J0513.1$+$0851\tablenotemark{b} & 60$\pm$10 & 30$\pm$5 & 0.067 & 0.44 & 4.018 \\
EK Cep\tablenotemark{c} & 23$\pm$2 & 10.5$\pm$2.0 & 0.109 & 0.55 & 4.43 \\
Parenago 2494\tablenotemark{d} & 22.0$\pm$2.5 & 11.4$\pm$2.0 & 0.257 & 0.71 & 19.48 \\
RX J0441.0-0839\tablenotemark{e} & 2$\pm$1 & 1$\pm$1 & 0.216 & 0.82 & 13.56 \\
RX J0350.5-1355\tablenotemark{e} & 19$\pm$2 & 7$\pm$1 & 0.0 & 0.92 & 9.28 \\

\enddata

\tablenotetext{a}{This paper}
\tablenotetext{b}{Ru\'{i}z-Rodr\'{i}guez et al. (2013)}
\tablenotetext{c}{$L$andin et al. (2009)}
\tablenotetext{d}{Reipurth et al. (2002)}
\tablenotetext{e}{Covino et al. (2001)}

\end{deluxetable}

\newpage

\begin{figure}
\centering
\includegraphics{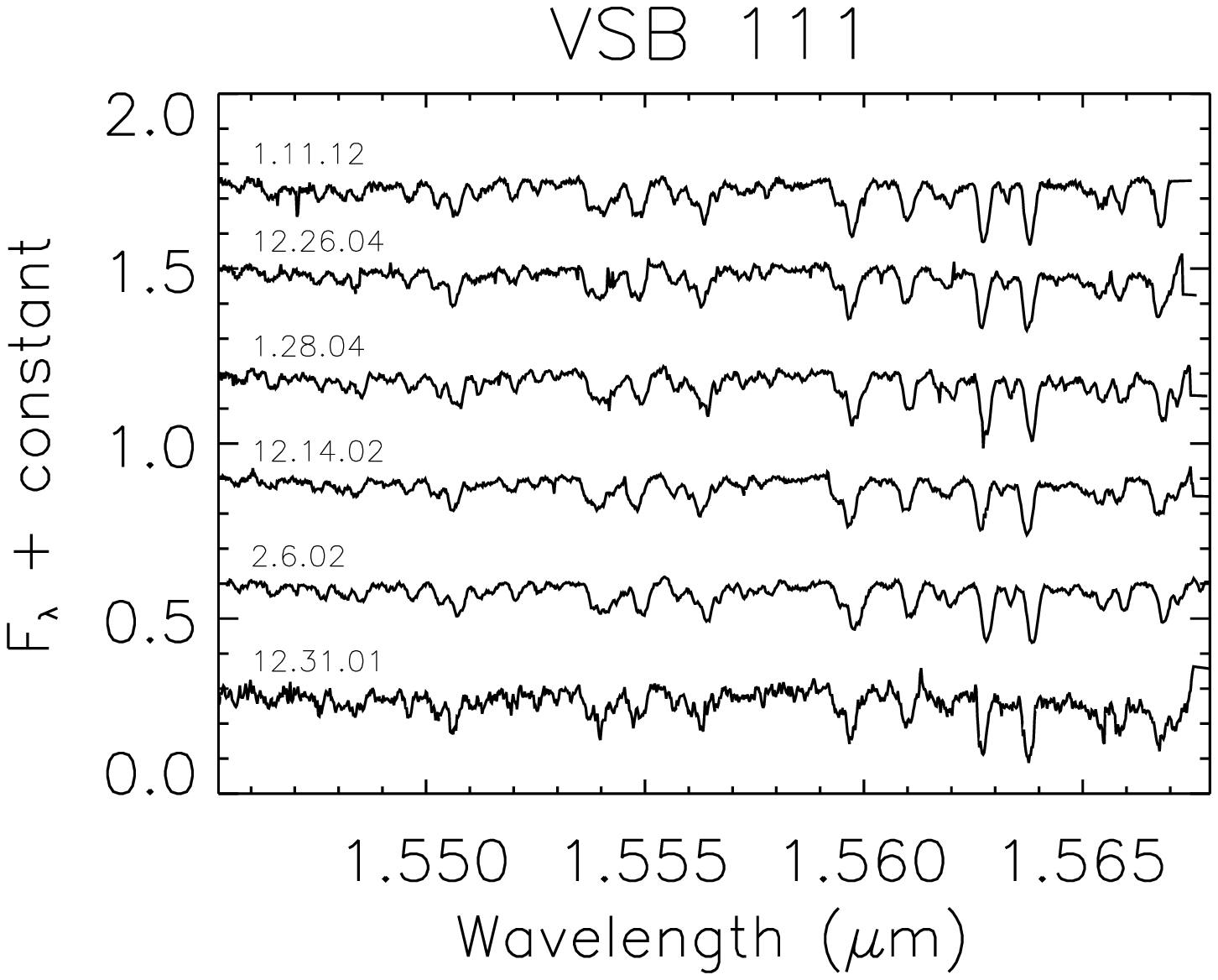}
\caption{Seven reduced spectra for VSB 111 with the correction for heliocentric motion taken into account.  UT dates are specified with corresponding spectra.}
\label{fig:fig1}
\end{figure}

\newpage

\begin{figure}
\begin{center}
\includegraphics{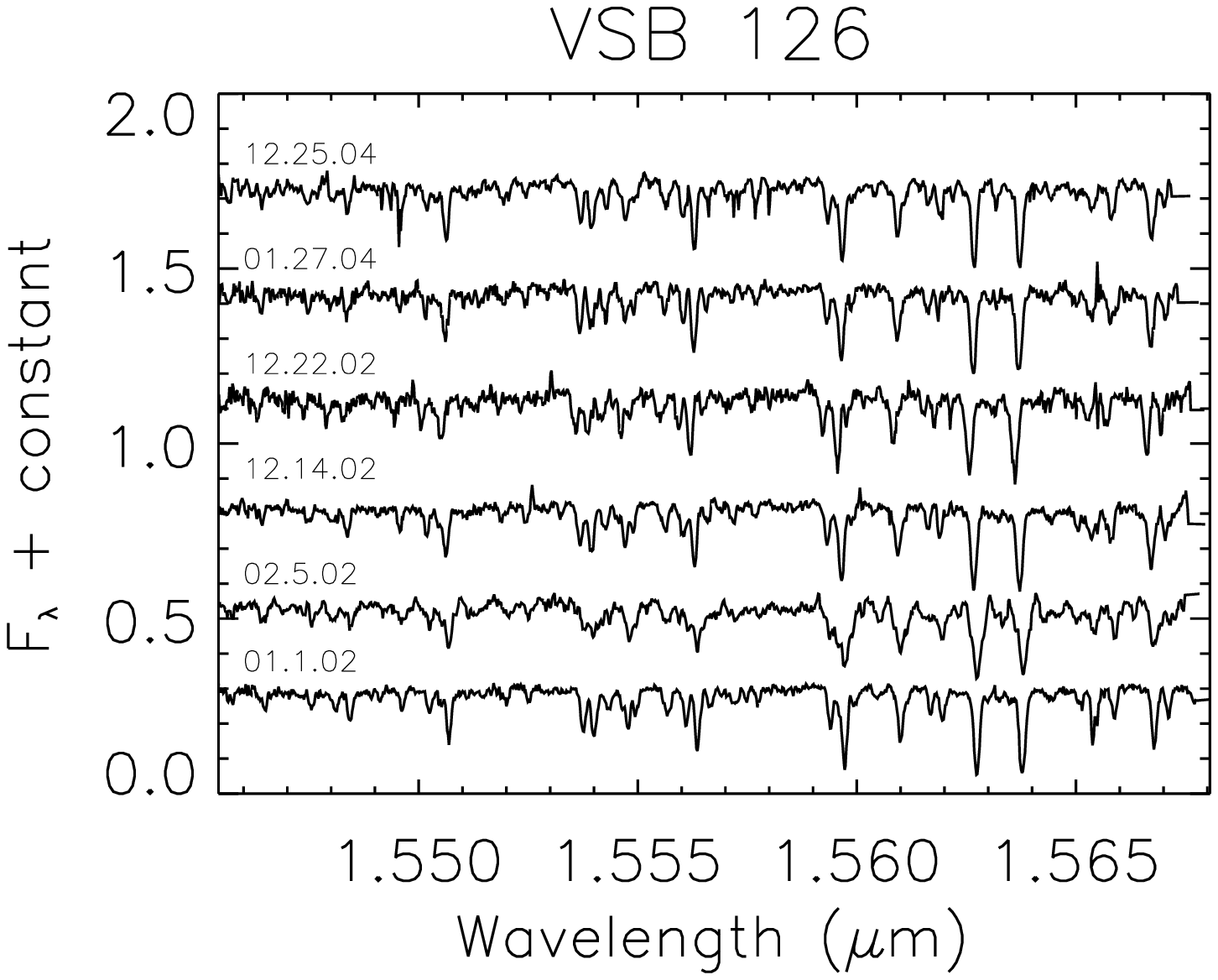}
\caption{Six reduced spectra for VSB 126 with the correction for heliocentric motion taken into account.  UT dates are specified with corresponding spectra.} 
\end{center}

\end{figure}
\label{fig:fig2}

\newpage

\begin{figure}
\begin{center}
\includegraphics{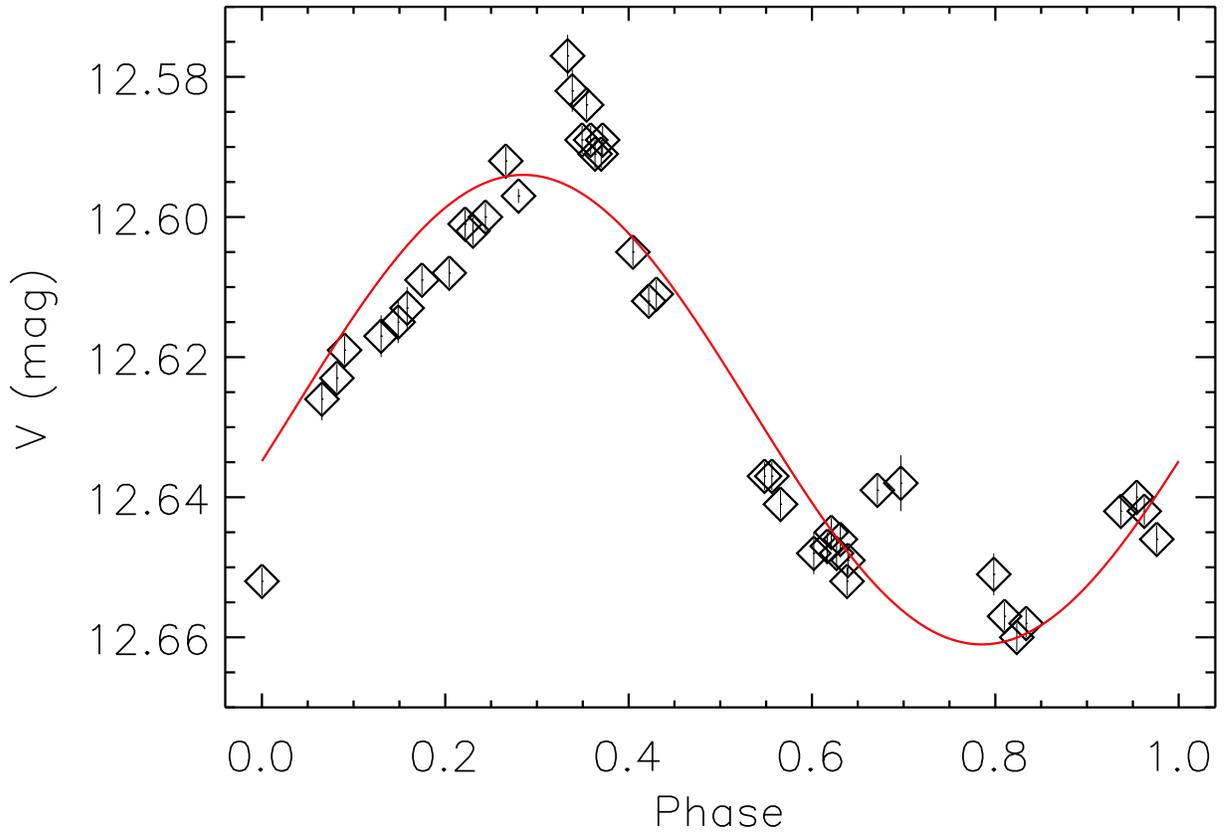}
\caption{Magnitude versus phase for the primary component of VSB 111. The rotation period was determined to be 3.74$\pm$0.02 days. The magnitudes are binned
in JD date to minimize the errors, shown as vertical bars. The red line is a weighted sinusoidal fit. The first image taken was used to phase the light curve and represents phase zero.}
\end{center}

\end{figure}
\label{fig:fig3}
	
\newpage

\begin{figure}
\begin{center}
\includegraphics{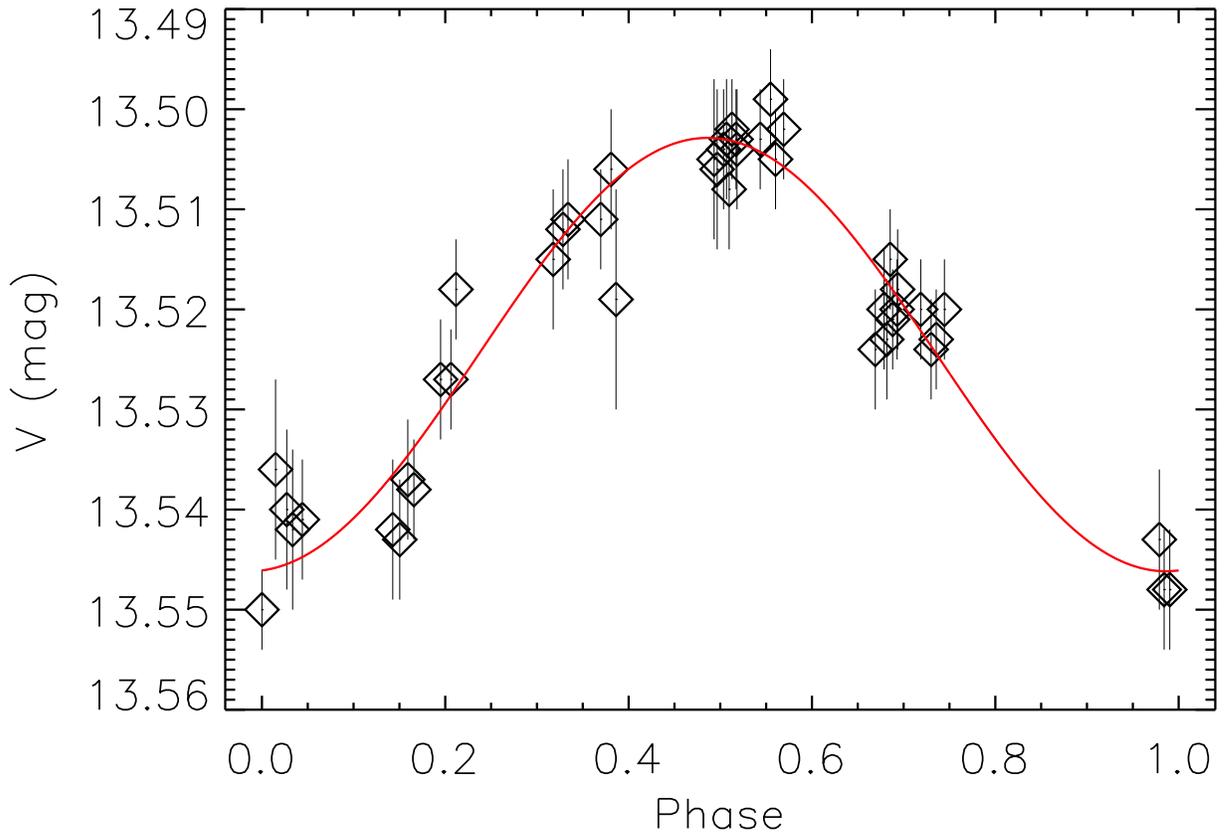}
\caption{Same as Figure 3 for VSB 126. The rotation period was determined to be 5.71$\pm$0.07 days.}
\end{center}

\end{figure}
\label{fig:4}

\newpage

\begin{figure}
\includegraphics{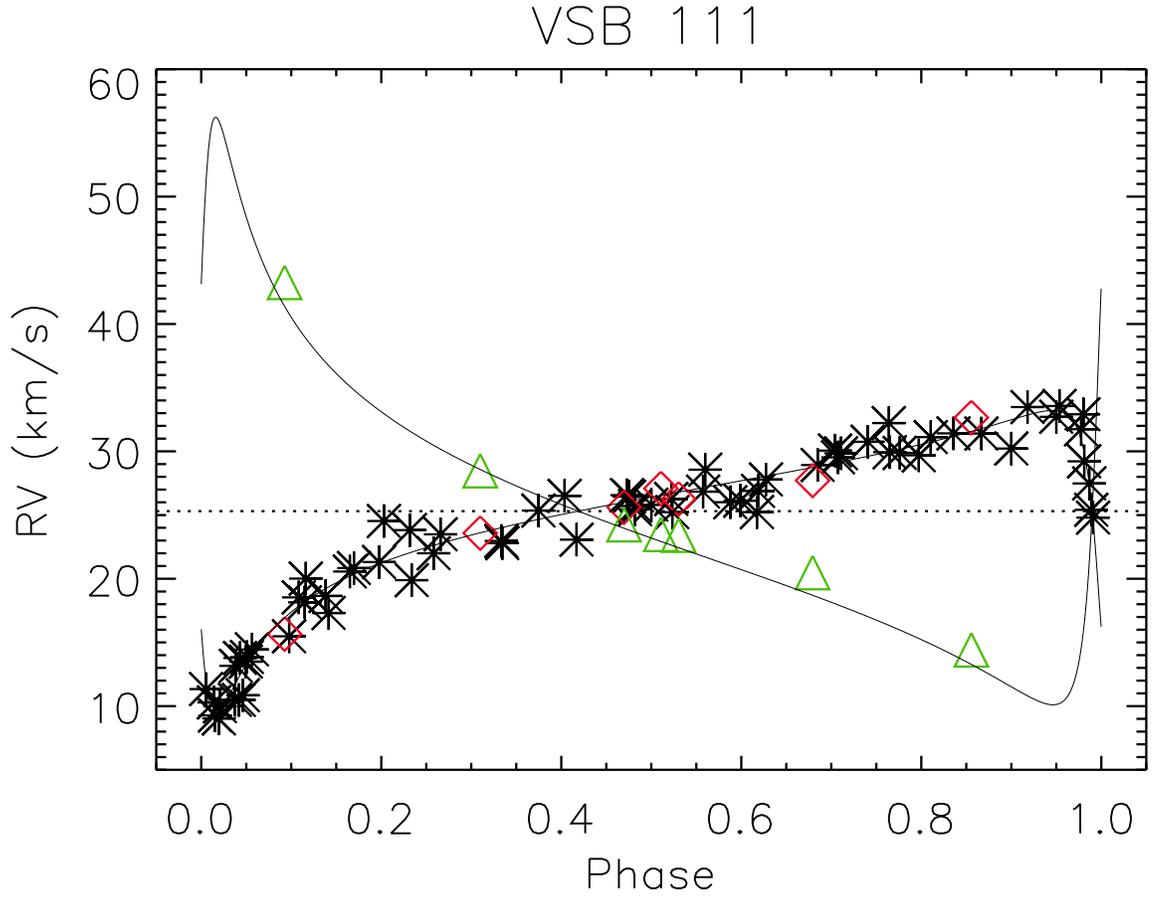}
\caption{Radial velocity versus phase for VSB 111. Measured RVs for the primary in the optical are represented by black asterisks and in the IR by red diamonds. Measured RVs in the IR for the secondary are represented by green triangles. The dotted horizontal line is the center-of-mass velocity, $\gamma$.  RV uncertainties are smaller than the plot symbol
size and are given in Tables 2 and 4.}
\begin{center}
\end{center}

\end{figure}
\label{fig:fig5}

\newpage

\begin{figure}
\begin{center}
\includegraphics{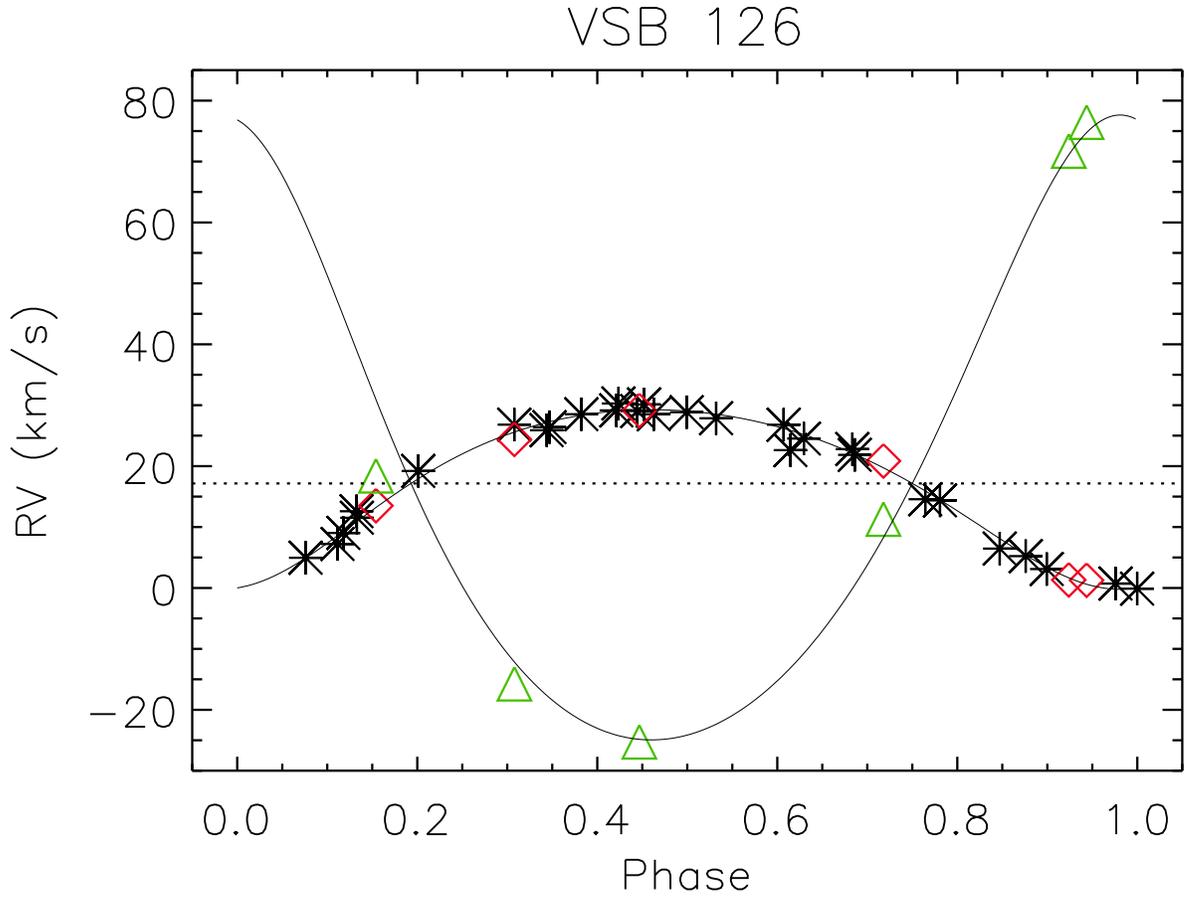}
\caption{Same as Figure 5 for VSB 126.  RV uncertainties are smaller than the plot symbol
size and are given in Tables 3 and 5.}
\end{center}

\end{figure}
\label{fig:fig6}

\newpage

\begin{figure}
\begin{center}
\includegraphics{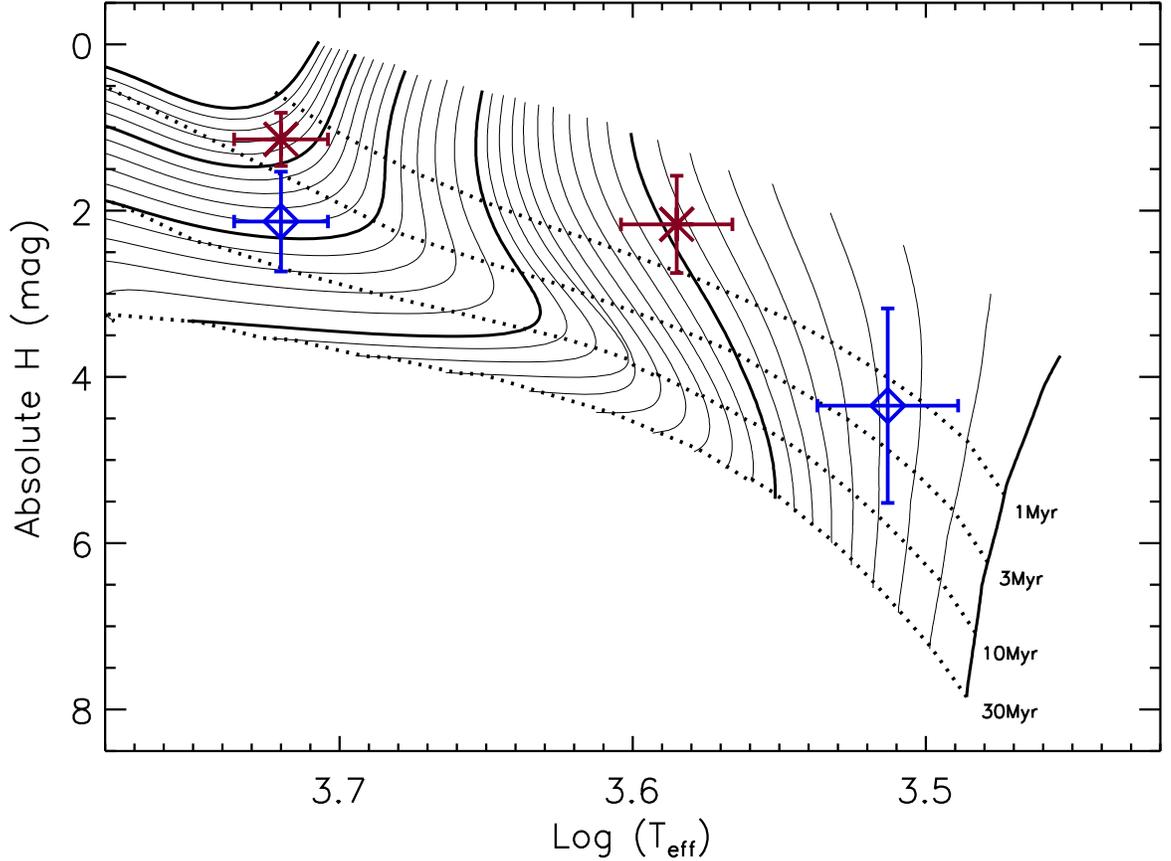}
\caption{Absolute H magnitude and Log($T_{\rm eff}$) for components of VSB 111 (red asterisks) and VSB 126 (blue diamonds) plotted on the PMS evolutionary tracks from the Dartmouth Stellar Evolution Database for solar metallicity = 0.0, ($\alpha$/Fe) = 0.0, and mixing length = 1.938. The absolute $H$ magnitudes, adopting a distance of 800 pc (Walker 1956), for the components were derived from 2MASS magnitudes and the spectroscopic flux ratios. The isochrones are labeled. The mass tracks start at 0.10M$_\odot$ and go to 1.0M$_\odot$ at 0.05M$_\odot$ intervals then the interval is 0.10M$_\odot$ from 1.0M$_\odot$ to 2.5M$_\odot$. The bold mass tracks represent 0.1M$_\odot$, 0.5M$_\odot$, 1.0M$_\odot$, 1.5M$_\odot$, 2.0M$_\odot$, and 2.5M$_\odot$. The errors in absolute magnitude are most sensitive to distance.} 
\end{center}

\end{figure}
\label{fig:fig7}

\newpage

\begin{figure}
\begin{center}
\plotone{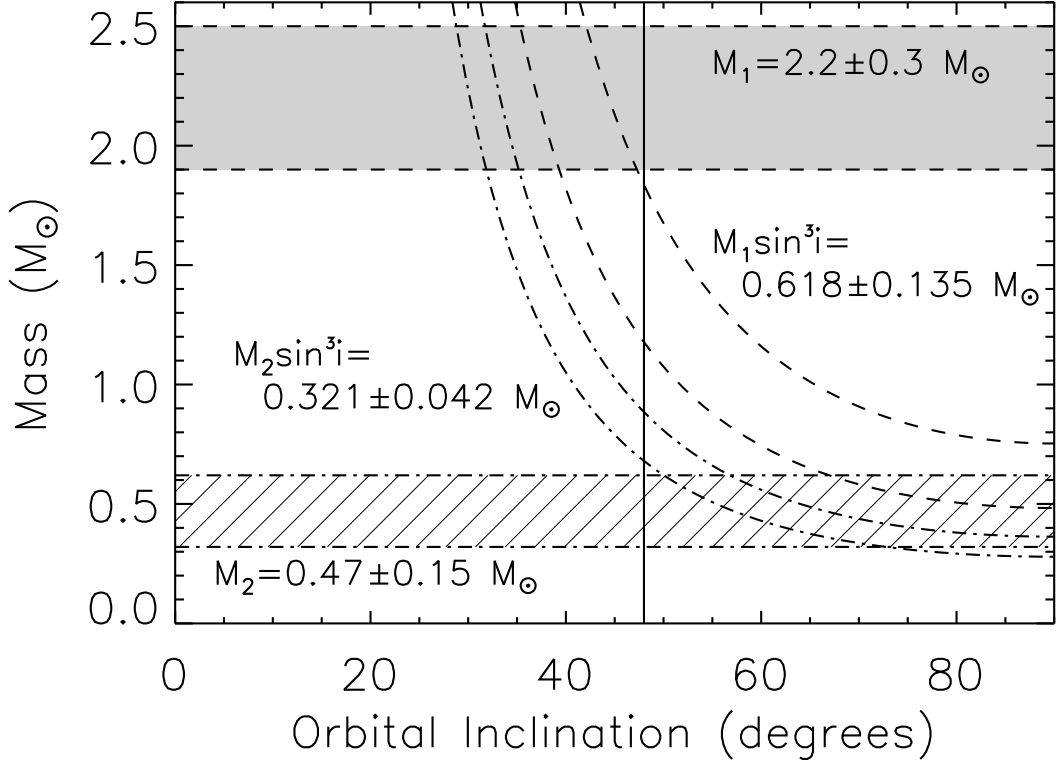}
\caption{The component masses of VSB 111 as a function of orbital inclination. The values of $M_1 \sin^3 i$ and $M_2 \sin^3 i$ are given from the orbital solution represented by the curved dashed and dash-dotted lines, respectively.
The shaded area represents the primary mass $\pm$1$\sigma$ and the hashed area represents the secondary mass $\pm$1$\sigma$, both derived from the Dotter et al. (2008) tracks. The vertical line shows approximate agreement between all parameters for an orbital inclination of $\sim$48$\,^{\circ}$.}
\end{center}

\end{figure}
\label{fig:fig8}

\newpage

\begin{figure}
\begin{center}
\includegraphics{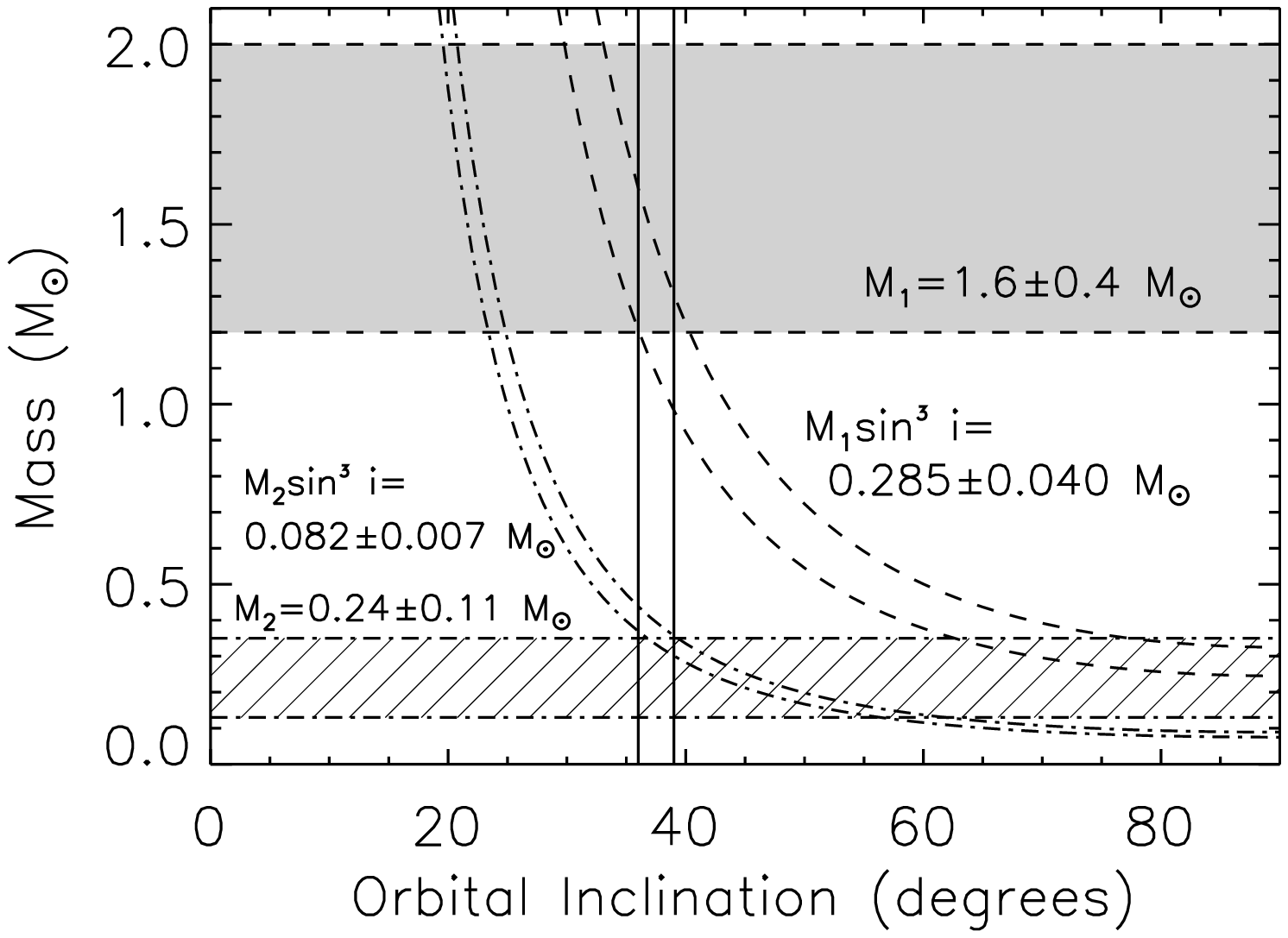}
\caption{Same as Figure 8 for VSB 126.  The vertical lines show agreement between all parameters for an orbital inclination of $\sim36-39 \,^{\circ}$.}
\end{center}

\end{figure}
\label{fig:fig9}







\newpage

\begin{thebibliography}{1}\kern\bibindent
 \bibliographystyle{apj}

\bibitem[Baraffe et al. (1998)]{bar98} Baraffe, I., Chabrier, G., Allard, F., \& Hauschildt, P.~H. \ 1998, A\&A, 337, 403

\bibitem[Bate (2009)]{bat09} Bate, M.~R. \ 2009, MNRAS, 397, 232 



\bibitem[Boyajian et al. (2012)]{boy12} Boyajian, T.~S., von Braun, K., van Belle, G., McAlister, H.~A., ten Brummelaar, T.~A., Kane, S.~R., Muirhead, P.~S., Jones, J., White, R., Schaefer, G., Ciardi, D., Henry, T., L{\'o}pez-Morales, M., Ridgway, S., Gies, D., Jao, W.~C, Rojas-Ayala, B., Parks, J.~R., Sturmann, L., Sturmann, J., Turner, N.~H., Farrington, C., Goldfinger, P.~J., \& Berger, D.~H. \ 2012, \apj, 757, 112

\bibitem[Buie(2010)]{bui10} Buie, M.~W.\ 2010, Advances in Astronomy, 2010, 130172


\bibitem[Charbonneau(1995)]{1995ApJS..101..309C} Charbonneau, P.\ 1995, \apjs, 101, 309

\bibitem[Covino et al.(2001)]{cov01} Covino, E., Melo, C., Alcal\'{a}, J.~M., Torres, G., Fern\'{a}ndez, M., Frasca, A., \& Paladino, R. \ 2001, A\&A, 375, 130


 
\bibitem[Dahm et al.(2007)]{dah07} Dahm, S.~E., Simon, T., Proszkow, E.~M., \& Patten, B.~M., 2007, \aj, 134, 999 

\bibitem[Dahm(2008)]{dah08} Dahm, S.~E., 2008, \ Handbook of Star Forming Regions, Volume I, 966

\bibitem[de Bruijne(2012)]{2012Ap&SS.341...31D} de Bruijne, J.~H.~J.\ 2012, \apss, 341, 31

\bibitem[Dotter et al. (2008)]{dot08} Dotter, A., Chaboyer, B., Jevremovi\'{c}, D., Kostov, V., Baron, E., \& Ferguson, J.~W. \ 2008, ApJS, 178, 89

\bibitem[Feldbrugge \& van Genderen(1991)]{fel91} Feldbrugge, P.~T.~M. \& van Genderen, A.~M., 1991, A\&A, 91, 209

\bibitem[Flaccomio et al.(1999)]{fla99} Flaccomio, E., Micela, G., Sciortino, S., Favata, F., Corbally, C., \& Tomaney, A., \ 1999, A\&A, 345, 521 

\bibitem[Flaccomio et al.(2006)]{fla06} Flaccomio, E., Micela, G., \& Sciortino, S.,  2006, A\&A, 455, 903

\bibitem[F\H{u}r\'{e}sz et al.(2006)]{fur06} F\H{u}r\'{e}sz, G., Hartmann, L.~W., Szentgyorgyi, A.~H., Ridge, N.~A., Rebull, L., Stauffer, J., Latham, D.~W., Conroy, M.~A., Fabricant, D.~G., \& Roll, J. 2006, \aj, 648, 1090

\bibitem[Goldberg et al.(2002)]{gol02} Goldberg, D., Mazeh, 
T., Latham, D.~W., Stefanik, R.~P., Carney, B.~W., \& Laird, J.~B. \ 2002, \aj, 124, 1132

\bibitem[Haisch et al.(2001)]{hai01} Haisch, K.~E., Lada, E.~A., \& Lada, C.~L. \ 2001, \apj, 553, L153

\bibitem[Hillenbrand (1997)]{hil97} Hillenbrand, L.~A. \ 1997, \aj, 113, 1733

\bibitem[Hillenbrand \& White (2004)]{hilwhi04} Hillenbrand, L.~A., \& White, R.~J. \ 2004, \apj, 604, 741



\bibitem[Kearns \& Herbst(1998)]{kea98} Kearns, K.~E., \& Herbst, W. \ 1998, \aj, 116, 261

\bibitem[Landin et al.(2009)]{lan09} Landin, N.~R., Mendes, L.~T.~S., \& Vaz, L.~P.~R. \ 2009, A\&A, 494, 209L

\bibitem[Latham(1992)]{lat92} Latham, D.~W. 1992, in Astronomical Society of the Pacific Conference Series, Vol. 32, IAU Colloq. 135: Complementary Approaches to Double and Multiple Star Research, ed. H.~A. McAlister \& W.~I. Hartkopf, 596

\bibitem[Latham et al.(2002)]{lat02} Latham, D.~W., Stefanik, R.~P., Torres, G., Davis, R.~J., Mazeh, T., Carney, B.~W., Laird, J.~B., \& Morse, J.~A. \ 2002, \aj, 124, 1144

\bibitem[Luhman (1999)]{luh99} Luhman, K.~L. \ 1999, \apj, 525, 466

\bibitem[Luhman et al.(2003)]{luh03} Luhman, K.~L., Brice\~{n}o, C., Stauffer,  J.~R., Hartmann, L., Barrado Y Navascu\'{e}s, D., \& Caldwell, N. \ 2003, \apj, 590, 348

\bibitem[Mace et al.(2009)]{mac09} Mace, G.~N., Prato, L., Wasserman, L.~H., Schaefer, G.~H., Franz, O.~G., \& Simon, M. \ 2009, \aj, 137, 3487

\bibitem[Mace et al.(2012)]{mac12} Mace, G.~N., Prato, L., Torres, G., Wasserman, L.~H., Mathieu, R.~D., \& McLean, I.~S. 2012, \aj, 144, 55

\bibitem[Makidon et al.(2004)]{mak04} Makidon, R. B., Rebull, L. M., Strom, S. E., Adams, M. T., \& Patten, B. M. \ 2004, \aj, 127, 2228

\bibitem[Mathieu(1994)]{mat94} Mathieu, R. \ 1994, A\&A, 32, 465

\bibitem[Mazeh et al.(2002)]{maz02} Mazeh, T., Prato, L., Simon, M., Goldberg, E., Norman, D., \& Zucker, S. \ 2002, \apj, 564, 1007

\bibitem[Mazeh et al.(2003)]{maz03} Mazeh, T., Simon, M., Prato, L., Markus, B., \& Zucker, S. \ 2003, \apj, 599, 1344

\bibitem[McLean et al.(1998)]{mcl98} McLean, I.~S., Becklin, E.~E. \& Bendiksen, O. et al. \ 1998, Proc. SPIE, 3354, 566

\bibitem[McLean et al.(2000)]{mcl00} McLean, I.~S., Graham, J.~R., \& Becklin, E.~E. et al. \ 2000, Proc. SPIE, 4008, 1048


\bibitem[Mendoza \& G\'{o}mez(1980)]{men80} Mendoza, E.~E \& G\'{o}mez, T. \ 1980, MNRAS, 190, 623


\bibitem[Nordstr\"om et al.(1994)]{nor94} Nordstr\"om, B., Latham, D.~W., Morse, J.~A., \ 1994, \aap, 287, 338


\bibitem[P\'{e}rez et al.(1987)]{per87} P\'{e}rez, M. R., Th\'{e}, P.~S., \& Westerlund, B.~E. 1987, \ PASP, 99, 1050

\bibitem[Prato(1998)]{pra98} Prato, L., 1998, Ph.D. thesis, SUNY Stony Brook

\bibitem[Prato et al.(2001)]{pra01} Prato, L., Ghez, A.~M., Pi\~{n}a, R.~K., Telesco, C.~M., Fisher, R.~S., Wizinowich, P., Lai, O., Acton, D.~S., \& Stomski, P. \ 2001, \apj, 549, 590

\bibitem[Prato et al.(2002)]{pra02} Prato, L., Simon, M.,  Mazeh, T., McLean, I.~S., Norman, D., \& Zucker, S. 2002, \apj, 569, 863


\bibitem[Prato et al.(2003)]{pra03} Prato, L., Greene, T.~P., \& Simon, M. 2003, \apj, 584, 853

\bibitem[Press et al.(1992)]{pre92} Press, W.~H., Teukolsky, S.~A., Vetterling, W.~T., \& Flannery, B.~P. \ 1992, Numerical Recipes in Fortran: The Art of Scientific Computing (2nd ed.; Cambridge: Cambridge Univ. Press) 

\bibitem[Ramirez et al.(2004)]{ram04} Ramirez, S.~V., Rebull, L., Stauffer, J., Hearty, T., Hillenbrand, L.~A. 2004, \aj, 127, 2659

\bibitem[Rebull et al.(2002)]{reb02} Rebull, L.~M., Makidon, R.~B., Strom, S.~E., Hillenbrand, L.~A., Birmingham, A., Patten, B.~M., Jones, B.~F., Yagi, H., \& Adams, M.~T. \ 2002, \aj, 123, 1528

\bibitem[Reipurth et al.(2002)]{rei02} Reipurth, B., Lindgren, H., Mayor, M., Mermilliod, J., \& Cramer, N. \ 2002, \aj, 124, 2813

\bibitem[Rojas-Ayala et al. (2012)]{roj12} Rojas-Ayala, B., Covey, K.~R., Muirhead, P.~S., \& Lloyd, J.~P. \ 2012, \apj, 748, 93


\bibitem[Rousselot et al.(2000)]{rou00} Rousselot, P., Lidman, C., Cuby, J.-G., Moreels, G., \& Monnet, G.\ 2000, A\&A, 354, 1134

\bibitem[Ru\'{i}z-Rodr\'{i}guez et al.(2013)]{rui13} Ru\'{i}z-Rodr\'{i}guez, D., Prato, L., Torres, G., Wasserman, L., \&  Neuh\"auser, R. \ 2013, \aj, 145, 162

\bibitem[Sagar \& Joshi (1983)]{sag83} Sagar, R. \& Joshi, U.~C. \ 1983, MNRAS, 205, 747 

\bibitem[Samus et al.(2012)]{sam12} Samus, N.~N., Durlevich, O.~V., Kazarovets, E.~V., Kireeva, N.~N., Pastukhova, E.~N., \& Zharova, A.~V.\ 2012,\ General Catalog of Variable Stars


\bibitem[Schaefer et al.(2012)]{sch12} Schaefer, G.~H., Prato, L., Simon, M., \& Zavala, R.~T. \ 2012, \apj, 756, 120

\bibitem[Siess et al. (2000)]{sie00} Siess, L., Dufour, E., \& Forestini, M. \ 2000, A\&A, 358, 593


\bibitem[Simon(2008)]{sim08} Simon, M.\ 2008, The Power of  Optical/IR Interferometry: Recent Scientific Results and 2nd Generation,  227

\bibitem[Sung et al.(1997)]{sun97} Sung, H., Bessell, M.~S., \& Lee, S.-W. \ 1997, \aj, 114, 2644

\bibitem[Sung et al.(2004)]{sun04} Sung, H., Bessell, M.~S., \& Chun, M.~Y. \ 2004, \aj, 128, 1684

\bibitem[Tognelli et al.(2011)]{tog11} Tognelli, E., Prada Moroni, P.~G., \& Degl' Innocenti, S. \ 2011, A\&A, 533, A109

\bibitem[Tokunaga(2000)]{tok00} Tokunaga, A.~T. 2000, Allen's Astrophysical Quantities, ed. 4, A. N. Cox (New York: Springer-Verlag), 143

\bibitem[Torres et al.(1997)]{tor97} Torres, G., Stefanik, R.~P., \& Latham, D.~W. 1997, \aj, 474, 256

\bibitem[Torres et al.(2002)]{tor02} Torres, G., Neuh\"auser, R., \& Guenther, E.~W. 2002, \aj, 123, 1701

\bibitem[Torres et al. (2013)]{tor13} Torres G., Ru\'{i}z-Rodr\'{i}guez, D., Badenas, M., Prato, L., Schaefer, G.~H., Wasserman, L.~H., Mathieu, R.~D., \& Latham, D.~W. \ 2013, \apj, 698, 242

\bibitem[Vasilevskis et al.(1965)]{vas65} Vasilevskis, S., Sanders, W. ~L., \& Balz, Jr, A.~G.~A. \ 1965, \aj, 70, 10, 797


\bibitem[Walker(1956)]{wal56} Walker, M.~F. 1956, \apj, Supplement Series, 2, 365

\bibitem[Warner(2007)]{war07} Warner, B.~D. 2007, Minor Planet Bulletin, 34, 113

\bibitem[Warner(2011)]{war11} Warner, B.~D. 2011, MPO Software, Canopus version, 10.4.0.6 

\bibitem[White et al. (1999)]{whi99} White, R.~J., Ghez, A.~M., Reid, I.~N., \& Schultz, G. \ 1999, \apj, 520, 811


\bibitem[Zucker \& Mazeh(1994)]{zuc94} Zucker, S., \& Mazeh, T. 1994, \apj, 420, 806

\end{thebibliography}
\end{document}